# Bead-Droplet Reactor for High-Fidelity Solid-Phase Enzymatic DNA Synthesis

Punnag Padhy[1*], Mohammad Asif Zaman[1], Michael Anthony Jensen[2,3*], Yao-Te Cheng[1], Yogi Huang[4], Ludwig Galambos[1], Ronald Wayne Davis[2,3,5], Lambertus Hesselink[1*]

[1]Department of Electrical Engineering, Stanford University, Stanford, CA, 94305, U.S.A
[2]Stanford Genome Technology Center, Stanford University, Palo Alto, CA, 94304, U.S.A
[3]Department of Biochemistry, Stanford University, Stanford, CA, 94305, U.S.A
[4]Department of Chemical Engineering, Stanford University, Stanford, CA, 94305, U.S.A
[5]Department of Genetics, Stanford University, Stanford, CA, 94305, U.S.A

*Corresponding author. Email: punnag@stanford.edu, m.a.jensen@stanford.edu, hesselink@ee.stanford.edu

**Solid-phase synthesis techniques underpin the synthesis of DNA, oligopeptides, oligosaccharides, and combinatorial libraries for drug discovery. State-of-the-art solid-phase synthesizers can produce oligonucleotides up to 200-300 nucleotides while using excess reagents. Accumulated errors over multiple reaction cycles prevent the synthesis of longer oligonucleotides for the genome scale engineering of synthetic biological systems. The sources of these errors in synthesis columns are poorly understood. Here we show that bead-bead stacking significantly contributes to reaction errors in columns by analyzing enzymatic coupling of fluorescently labelled nucleotides onto the initiated beads along with porosity, particle tracking and diffusion calculations. To circumvent stacking, we introduce dielectrophoretic bead-droplet reactor (DBDR); a novel approach to synthesize on individual microbeads within microdroplets. Dielectrophoretic force overcomes the droplet-medium interfacial tension to encapsulate and eject individual beads from microdroplets in a droplet microfluidic device. Faster reagent diffusion in droplets, and non-uniform electric field induced enhancement in reagent concentration at its surface can improve reaction fidelities in DBDR. Fluorescence comparisons suggest around 3-fold enhancement of reaction fidelity compared to columns. DBDR can potentially enable the high-purity synthesis of arbitrarily long strands of DNA to meet the emerging demands in healthcare, environment, agriculture, materials, and computing.**

# 1. Introduction

Solid-phase synthesis techniques [1] revolutionized de novo synthesis of DNA [2-4], oligopeptides [1], oligosaccharides [5], and combinatorial libraries for drug discovery [6]. Insoluble solid support bound molecules can be sequentially exposed to a large excess of reagents to synthesize the desired long-chain molecule step-by-step [1]. Current state-of-the-art platforms can synthesize oligonucleotides (single-stranded DNA) that are 200-300 nucleotides long in practice [2-4, 7-9], irrespective of the synthesis chemistry. Accumulated errors over multiple reaction cycles limit the synthesis of longer oligonucleotide strands in the existing solid-phase synthesis platforms [2-4, 7-9]. For example, in enzymatic synthesis [7, 10-12], these errors lead to a large polydispersity in the synthesized oligonucleotides and reduce the yield of the desired product with increase in intended strand length [12]. However, as synthetic biology seeks to engineer increasingly complex biological systems, there is a surge in demand for the synthesis of arbitrarily long strands of oligonucleotides [8, 9]. This requires the minimization of errors over individual synthesis reaction steps. Therefore, development of fundamentally novel physical sample handling techniques for solid-phase synthesis that can facilitate more efficient interfacing of solid supports and reagents to enable high-purity synthesis reactions is essential. Faster reaction rates, reduced reagent consumption and wastage as well as a cleaner synthesis process are additionally desirable [8].

Development of high-purity synthesis techniques necessitates an understanding of the physical basis of inadequate reagent exposure in existing platforms. In microarrays, misalignment of optical beams and inkjet print heads leads to substitution and deletion errors [2-4]. In columns, which tend to be more efficient and amenable for subsequent gene assembly [2-4], deletion strands are the major byproduct. However, the actual reason for the errors is poorly understood and generally attributed to the reaction chemistries that are less than 100% efficient [8]. As a result, significant research is directed towards optimization and improvement of the chemistries [2, 9-11]. Research on physical sample handling is primarily geared towards enhancing parallelization, throughput, and miniaturization [7, 8].

We establish that the stacking of beads in synthesis columns (Figure 1a) curtails adequate exposure of the growing oligonucleotide strands on the bead surfaces to the reagents and limits the reaction fidelity. To address this problem, we postulate that synthesis on individual beads can maximize reagent access to the bead surfaces. To ensure adequate reagent usage and minimum wastage per bead we choose to synthesize within microdroplets. Furthermore, as an added benefit, the small size of the microdroplet can facilitate faster reagent diffusion [13-15] to the reaction site to drive faster reactions. The above objectives can be achieved by encapsulating and ejecting individual beads (of radius $R_b$) from microdroplets (of radius $R_d$). However, capillary forces due to interfacial tension [13, 16] ($\gamma_{ow}$) are significant at this length scale and must be overcome to manipulate the bead across the droplet-medium interface. In state-of-the-art bead encapsulation techniques, beads pre-suspended in the dispersed (droplet forming) phase are flowed towards the droplet generator site [17, 18]. After the droplet breakup, the bead is held inside it by the droplet-medium interfacial tension. It is not amenable to implement multistep solid-phase synthesis reactions on individual beads using separate reagent microdroplets for each step. Hence, there is a need for new approaches to encapsulate the bead from the surrounding immiscible medium into the droplet and eject it back out into the medium.

Herein, we introduce DBDR (Dielectrophoretic Bead-Droplet Reactor), a technique to implement solid-phase oligonucleotide synthesis on individual beads by using dielectrophoretic force [19-23] to encapsulate and eject them from picoliter scale reagent microdroplets (Figure 1b). When sufficiently large voltage ($V_s$) is applied across the electrodes, the dielectrophoretic force ($\propto R_d^3 |V_s|^2$) overcomes the capillary forces ($\propto R_b \gamma_{ow}$) to move the hydrophobic bead, across the aqueous reagent-oil medium interface, into the droplet. At low voltages ($V_s$), the dominant capillary force ejects the hydrophobic bead out of the droplet. As a proof-of-concept, we demonstrated the enzymatic coupling of fluorescently labelled nucleotides to the 3' end of the initiator strands on the bead using DBDR (Figure 1b). Furthermore, fluorescence measurements show that DBDR achieves higher fidelity of reactions than columns with stacked beads.

## 2. Device And Reaction Design

The silicon-on-glass microfluidic device developed for demonstration of DBDR consists of Indium Tin Oxide (ITO) electrodes ($\approx 15 \ \mu m$ wide, Figure 1c and d) in the reaction chamber that are suitably aligned with the on-demand droplet generator microchannel [16, 24] ($\approx 25 \ \mu m$ wide, Figure 1c and d) to ensure that the dispensed droplet ($R_d \approx 25 \ \mu m$) lies within its trapping range. The electrodes are made smaller than the channel (Figure 1d) to ensure that the trapped droplet encapsulates the trapped bead as it covers the electrode (Figure 1b). ITO and glass ensure optical transparency for imaging the trapping and manipulation process as well as fluorescence detection of the coupling reaction. (Supplementary Material and Supplementary Figure 1).

Streptavidin-coated fluorescent green polystyrene beads ($R_b = 3 \ \mu m$, Figure 1e) are used as solid supports. 5' biotinylated oligonucleotide with 25 nucleotides bound to the microbead using streptavidin-biotin linkages (Figure 1e(i)) are used as initiators. The enzyme terminal deoxynucleotidyl transferase (TdT) couples the fluorescently labelled nucleotides (dCTP-AF647) to the 3' end of the initiators (Figure 1e(ii)) when the bead is encapsulated within the reagent droplet (Figure 1b). A red LED excites AF647 to detect the coupling of dCTP-AF647 to the initiated bead (Figure 1b and 1e(ii), Supplementary Material and Supplementary Figure 2-4).

## 3. Bead-Droplet Encapsulation and Ejection

The device, filled with $\approx 2.5\%$ w/w solution of Span 80 in silicone oil (kinematic viscosity$= 1 \ cSt$), is mounted on the sample holder with relevant electrical and fluidic connections (Supplementary Figure 3a & b). Initiated beads suspended in the oil solution are introduced into the device through the bottom fluidic port (Figure 1d). When a single bead enters the vicinity of the electrodes it is dielectrophoretically trapped (Supplementary Figure 3c) by applying a voltage ($V_s$) across the electrodes ($E_1$ and GND in Figure 1d). Subsequently, a reagent droplet is dispensed into the reaction chamber (Supplementary Figure 3d) and trapped adjacent to the bead (Supplementary Figure 3e).

When $V_s$ is increased to $\approx 120\ V$ (amplitude), the bead is fully encapsulated by the reagent droplet (Figure 2a, Movie 1). Reducing $V_s$ to $\approx 0.1\ V$ ejects the bead from the droplet (Figure 2a, Movie S1). Electrohydrodynamic simulations (Figure 2b, Supplementary Material) support these observations. It can be understood in terms of the change in the system's electrocapillary potential energy ($\Delta U$) [25, 26] due to a change in the Gibbs free energy ($\Delta U_{IT} \approx -\gamma_{ow}\cos(\theta)\Delta A_{ws}$) of the interfaces between the bead, droplet and the medium as well as the change in the electrical energy ($\Delta U_E = -\Delta(QV_s)/2$) stored in the system (Figure 2c&d). Here, $\gamma_{ow} = 5.5 mN/m$, $A_{ws}$ is the area of the bead covered by the reagent droplet, $\theta = 145°$ is the contact angle that the reagent droplet forms on the surface of the bead (Figure 3) and $Q$ is the charge stored on the electrode $E_1$ (Supplementary Material). The large $\theta$ means the bead has a propensity towards the oil (hydrophobic). When $V_s$ is high, $|\Delta U_E| \gg |\Delta U_{IT}|$. To attain the minimum energy configuration, the droplet moves towards the electrode ($\Delta U \approx \Delta U_E < 0$) and engulfs the bead in the process (Figure 2c). When $V_s$ is low, $|\Delta U_E| \ll |\Delta U_{IT}|$. To attain the minimum energy configuration, the droplet ejects the hydrophobic bead, which is touching the droplet-medium interface from the inside, as it moves away from the electrode ($\Delta U \approx \Delta U_{IT} < 0$, Figure 2d). Once the bead completely separates from the droplet, $\Delta U_{IT}$ is negligible (Figure 2d). Design tradeoffs indicate that while a larger droplet would mean more reagent usage per reaction, a smaller droplet would require a larger voltage to exert significant dielectrophoretic force ($\propto R_d^3|V_s|^2$) to move the bead across the interface (Figure 3a). Our choice of droplet size ($R_d = 25\mu m$) is a good compromise. Silicone oil 1cSt acts as a chemically inert suspension medium with low viscosity (Figure 3b) and high dielectric breakdown strength that prevents the rapid evaporation of the tiny reagent microdroplets due to their high surface-to-volume ratio [27]. Addition of the surfactant (Span80) to silicone oil 1cSt further reduces $\gamma_{ow}$ to $5.5\ mN/m$ and increases $\theta$ to $145°$. This makes the otherwise hydrophilic streptavidin surface of the bead (water droplet forms a contact angle of $50°$ on a streptavidin surface in air) hydrophobic (Figure 3c). Therefore, the choice of silicone oil and surfactant is critical for the spontaneous ejection of the bead from the droplet on touching its interior surface in the absence of an applied voltage. Hence, we establish a robust approach to encapsulate and eject individual beads from droplets (detailed experimental process in Supplementary Material).

## 4. Enzymatic Coupling of Nucleotide In DBDR

After establishing the physical working principle of DBDR, it was used for a proof-of-concept demonstration of enzymatic coupling of single nucleotides onto the initiator strands bound to a microbead. The above process was duplicated along with fluorescent imaging of the bead with a red LED before encapsulating after ejecting it (Figure 4a, Supplementary Material and Supplementary Figure 4a-f). The clear red emission at the position of the bead after ejection confirmed the binding of the fluorescently tagged bases (dCTP-AF647) to the bead. Three repeats of this experiment (Supplementary Figure 4g) were performed with encapsulation time of 300s to ensure reproducibility. To eliminate any false positives due to unincorporated bases non-specifically bound to the surface of the bead, we reiterated the above process using beads devoid of initiator strands and reagent solution without TdT (control experiment in Figure 4b). Lack of red fluorescence at the site of the bead after ejection confirmed the absence of non-specific binding

(Figure 4b, Supplementary Materials and Supplementary Figure 4h-m). This establishes DBDR as a robust approach for enzymatic DNA synthesis on individual beads in droplet reactors. While prior reports of chemical synthesis in microdroplets involved all the reagents in fluidic form [28,29], there was no droplet microfluidic analogue to solid-phase synthesis [8, 30]. DBDR brings together the disparate fields of droplet microfluidics and solid-phase synthesis. While droplet microfluidic implementation of gene assembly, gene expression and other steps of the design-build-test cycle of synthetic biology exist, a droplet microfluidic implementation of solid-phase synthesis remained elusive due to the inability to handle solid supports in such systems [30]. By addressing this problem, DBDR opens a route for a completely droplet microfluidic process pipeline for synthetic biology that could minimize reagent consumption during testing and verification.

## 5. Coupling Fidelity Enhancement in DBDR

After having demonstrated enzymatic coupling of nucleotides to initiated beads using DBDR, we compared its coupling fidelity with reactions implemented in synthesis columns. Large variations in fluorescence intensity across beads in the column (Figure 5) indicate nonuniform reagent environment in the vicinity of each bead which results in corresponding variations in the reaction fidelities. Only a small fraction of the beads has extremely bright fluorescence which closely represent the maximum attainable enzymatic coupling fidelities. The reduced fidelity on an overwhelming majority of the beads leads to an overall decrease in the average coupling efficiency (average of the histogram in Figure 5) and suggest significant missed couplings. Beads reacted using DBDR consistently exhibit higher fluorescence intensities which is quantified using the average fluorescence intensity ratio ($FIR_{avg} = 3.2$, Figure 5). This indicates enhanced access of individual bead surfaces to reagents within the microdroplets which then translates to improved reaction fidelities. A 3.2 times higher reaction fidelity implies that reactions implemented using DBDR can achieve near-perfect fidelity even when the same reactions implemented in synthesis column attain a fidelity of mere 32%. While excess concentration of reagents is generally used to drive up reaction fidelities in synthesis columns [1, 31], it is accompanied by excess reagent wastage as well. If the synthesis processes are repeated to generate $N$ nucleotides long oligomer the ratio of error free products ($EFPR$) synthesized using DBDR and synthesis columns will be given by $EFPR = \frac{1}{10^6} \times 3.2^N$ (around a million beads are used in the synthesis column). Therefore, for merely $N = 12 > \frac{6}{\log_{10} 3.2}$, $EFPR > 1$ can be achieved. So, higher quantity of error free synthesis product can be achieved on 1 bead using DBDR than on a million beads using synthesis columns for oligomers that are longer than merely 12 nucleotides ($N = 12$). Furthermore, the ratio of reagents usage ($RUR$) between DBDR and synthesis columns would be $RUR = \frac{60 pl}{50 \mu l} \approx 10^{-6}$. Hence, DBDR has the potential to synthesize significantly higher quantity of error free strands while consuming much less reagents. The performance improvements would be more pronounced with increase in $N$. This observed superior performance of DBDR compared to synthesis columns (statistical significance established using Welch's t test, Supplementary Material) can be attributed to 1) bead-bead stacking, 2) faster reagent diffusion time in droplets compared to columns and 3) non-uniform electric field acting on droplet reactor.

Firstly, the stacking of beads in synthesis columns restricts reagent access to the bead surfaces. Ideally, beads can be stacked in a definite number of regularly packed lattice arrangements [32]. The rhombohedral [33, 34] is one of the most tightly packed arrangements with a packing fraction of 0.74. It has a fluid filled inter-particle void volume of $4\left(\sqrt{2} - \frac{\pi}{3}\right)R_b^3 \approx 40 fl$ [33, 34] in a single unit cell encompassing a bead (of radius $R_b$) with $\approx 140\ attomoles$ of initiators on the surface. On the other extreme, the simple cubic (SC) arrangement has a packing fraction of 0.52 and a void volume of $4\left(2 - \frac{\pi}{3}\right)R_b^3 \approx 103 fl$ [32, 33] in a single unit cell. The $5\mu M$ concentration of fluorescently labelled nucleotides implies $\approx 0.2\ attomoles$ ($\approx 140\ attomoles/700$) and $\approx 0.52 attomoles$ ($\approx 140\ attomoles/280$) of nucleotides per void in a unit cell in rhombohedral and SC stacking respectively. On the other hand, in DBDR, the bead has access to the entire reagent volume within the droplet (forms a contact angle $\varphi = 140°$ with the device surface, Supplementary Figure 7) which is $\frac{(2-3\cos\varphi+\cos^3\varphi)}{3}\pi R_d^3 = 65 pl$ [25] and $\approx 325\ attomoles$ ($\approx 140\ attomoles/0.43$) of nucleotides. As we can see, while the droplet has ample nucleotides to completely couple with the initiators on the bead surface, the voids in the stack don't. Therefore, ideally the fluorescence intensity ratio ($FIR_{avg}$) between beads reacted using DBDR and in perfectly stacked synthesis column is expected to be between 280-700. However, in practice perfect stacking is a rarity. On introduction of reagents into the synthesis column (say at $t = 0$, Figure 6a), the particles are displaced from the perfect stack due to the drag force [13] exerted by the turbulent flow [35] ($t = 0.05$, Figure 6a). The increased interparticle spacing facilitates better reagent access to the bead surfaces. This explains the much lower $FIR_{avg}$. Once the reagent influx stops, the particles settle down under gravity into imperfect stacks ($t = 70s$ and $t = 300s$, Figure 6a).

Different particles have different settling times. This explains the varying exposure (spread of the grey band and histogram in Figure 5) to reagents during the experimental time ($300s$). Even when the beads are floating up, there are many beads in proximity ($\approx 10$ beads in a region of size equivalent to a droplet in frame $t = 0.05$ in Figure 6a) competing for the reagent molecules. Therefore, we see that the tight stacking of beads in synthesis columns which is known to provide the large surface area of solid support for the synthesis of large quantity of oligonucleotides [2, 3] also contributes significantly to reaction errors (reduced reaction fidelity). These errors accumulate over multiple synthesis cycles to ultimately negate the advantage of large synthesis quantity achieved in the columns. On the other hand, by ensuring one bead per droplet, DBDR is uniquely placed to ensure maximum attainable reagent access (resulting in higher observed fluorescence) to the bead surface during the experimental duration. As a result, beads reacted using DBDR exhibit higher than maximum attainable fluorescence in synthesis columns. As stated earlier, to compensate for this reduced reagent access in synthesis columns, higher reagent concentrations are used [31]. However, this increases reagent wastage per reaction which can accumulate over multiple synthesis cycles to make the synthesis of longer strands inviable.

Secondly, in synthesis columns beads not only access reagents from the voids in their immediate vicinity but also molecules that diffuse to the stack from further reaches of the column to replenish the reacted molecules. The shorter dimensions of the droplet facilitate much faster diffusion ($T_{diff} \propto l^2$) of reagent species than columns [13-15]. For example, it would take a single nucleotides (with diffusion coefficient $D \approx 10^{-6} - 10^{-5} cm^2/s$ [36, 37]) approximately $T_{diff,col} = \frac{h^2}{2D} \approx 25000s$ ($h = 2.28mm$) to diffuse from the top of the column to the stack of beads at the bottom (Figure 6b) which is much longer than the experimentally employed reaction times ($\approx 300s$). On the other hand, diffusion time across the droplet ($R_D = 25\mu m$) is a mere $T_{diff,DBDR} = \frac{2R_D^2}{D} \approx 12.5s$. Therefore, within an experimental time of $300s$ the coupling reaction in DBDR is less likely to have unreacted strands than in synthesis columns. This can enable the faster synthesis of longer strands. Hence, by using parallel manipulation techniques [38] DBDR synthesis can be extended to multiple beads undergoing reactions in separate droplets to achieve desired synthesis throughput along with high-fidelity and faster reactions. Again, although excess reagent concentration can compensate for the longer diffusion times in synthesis columns, it will also increase the reagent wastage. It is worth mentioning that in an ideal synthesis column scenario, beads would be tightly packed between the filters. In such a case, only the reagents within the voids will be accessible for coupling. This would further reduce reaction fidelity and the average fluorescence intensities of the beads reacted in columns ($FIR_{avg}$ will increase).

Thirdly, the applied electric field can affect rates of catalysis by lowering the activation energy required to perform a chemical reaction [39, 40]. Furthermore, the confinement provided by the droplet surface along with the non-uniform AC electric field driven drift of the charged species such as e.g. $Na^+$, $Co^{2+}$, $Mn^{2+}$, $Zn^{2+}$, and $Mg^{2+}$ (represented generally as $Me^+$) and the negatively charged dCTP-AF647 in the reagent can lead to their significantly enhanced concentrations at the droplet-medium interfaces at opposite ends of the droplet (Figure 6c). With electric drift coefficients ($\mu_{drift}$) $\approx 3 \times 10^{-4} cm^2/V.s - 6 \times 10^{-4} cm^2/V.s$ [13, 36], these charged reacting species can migrate a distance of $75\mu m$ ($> 2R_d$) under the influence of an average electric field of $10kV/cm$ withing a half period ($2.5ms$) of a $200\,Hz$ AC supply. The increased concentration of $Me^+$ over the negative half cycle will enhance their acceptance into terminal transferase (TdT). Its rate of polymerization is dependent on which ion is bound to the active site ($Mg^{2+}$ incorporates $dGTP$ and $dATP$; $Co^{2+}$ incorporates $dCTP$ and $dTTP$) [41]. Furthermore, the electric field may also help stabilize the metal ion-triphosphate interaction, thus enabling a more efficient nucleophilic attack by the 3'-OH group of the primer on the α-phosphate of the dCTP-AF647 [42]. Their improved incorporation into the enzyme and the enhanced concentration of dCTP-AF647 in the positive half of the AC cycle can lead to improved coupling fidelities. It is noteworthy that additional reagents need not be introduced into the droplet to achieve these concentration enhancements. Therefore, this does not entail any additional reagent consumption or wastage. Our predictions are in line with observations of reaction rate accelerations in microdroplet reactors [43, 44].

## 6. Conclusion

In summary, we introduce a novel technique (DBDR) to implement enzymatic oligonucleotide synthesis on individual beads by combining dielectrophoresis and droplet microfluidics. Fluorescence measurements and modelling suggest that by avoiding bead-bead stacking reaction fidelities using DBDR can exceed synthesis columns. The stacking of beads in synthesis columns provides a large surface area for the synthesis of large quantities of oligonucleotides [2, 3]. However, it also enhances the accumulation of reaction errors which ultimately negates the product quantity advantage over multiple reaction cycles to limit the synthesized strand length. On the other hand, the superior reaction fidelity of DBDR, achieved due to improved interfacing of solid supports with reagents, when replicated over multiple reaction cycles using scalable device architectures can potentially enable the synthesis of ultra-long strands of oligonucleotides ($\gg$ 300 bases) with unprecedent yields of the intended FLP while minimizing reagent consumption and wastage. It will mitigate the need for subsequent purification and error correction and reduce the cost of synthesis per base [8]. The lower reagent diffusion times across the droplet may lead to faster reactions which can be advantageous in the synthesis of long strands. The electrodes exerting the non-uniform electric field can be seen as on-site alignment features integrated onto the synthesis platform that maximize interaction between individual beads (solid supports) and reagent droplets. The electrode identifies the difference in the electrical properties of the droplet, bead, and the suspension to exert a dielectrophoretic force to enforce their alignment. This is in sharp contrast to microarrays where the reagent alignment apparatus is external to the synthesis surface [2-4, 9] (mechanical alignment of micromirrors in DMDs and nozzles in inkjet printheads). Reagents are dispensed or synthesis spots are activated based on preprogramed sequence of coordinates. No feedback from the synthesis surface is followed to ensure alignment accuracy. Furthermore, our research expanded the functionality of droplet microfluidics based chemical synthesis systems by incorporating solid-phase synthesis.

## 7. Outlook and Scope for Future Work

In addition to being a potential platform for high-fidelity solid-phase enzymatic DNA synthesis, DBDR can have many other applications. Water-in-oil droplets are excellent model systems to study enzymatic reactions within volume-confined environments [45, 46]. Typically, droplets are merged to deliver the reaction substrate to the enzyme. However, this process is accompanied by an increase in the overall volume. Instead, if the substrate is bound to a solid support, as in DBDR, the volume of the aqueous compartment and the reagent concentrations remains the same. Droplet microfluidics has enabled the processing of chemical information using Boolean logic, paving the way for chemical computation as well as logic-driven manipulation of chemical payloads for combinatorial screening applications [47]. DBDR can extend this concept a step further by writing chemical information onto solid supports; thus, creating solid-state drives for chemical information, which can facilitate one-bead-one-compound combinatorial synthesis [6, 48] using Boolean logic as well as DNA data storage on a droplet microfluidic processor. DBDR can be extended to implement other forms of solid-phase synthetic chemistries such as oligopeptide, and oligosaccharide. Finally, the diminishing costs of performing solid-phase combinatorial chemistry using the one-bead-one-compound [6, 48] approach on DBDR may make it economically feasible for a larger body of scientists to pursue research into traditionally expensive areas of chemistry such as drug discovery and vaccine development; this may expedite scientific discoveries to cure potentially life-threatening diseases and ailments.

**Acknowledgments:** We thank Paul Christopher Hansen for helpful discussions on the numerical simulations.

**Funding:** This project was funded by National Human Genome Research Institute (NHGRI) and National Institute of General Medical Sciences (NIGMS) of the National Institute of Health (NIH) under grant numbers 5R21HG009758 and R01GM138716 respectively and by the Stanford SystemX Alliance Seed Funding. Part of this work was performed at the Stanford Nano Shared Facilities (SNSF)/Stanford Nanofabrication Facility (SNF), supported by the National Science Foundation under award ECCS-2026822.

**Author contributions:** L.H., R.W.D., P.P., M.A.Z. and M.A.J. conceived the project. L.H. supervised the project. P.P. designed and fabricated the device with process inputs from M.A.Z., Y.C. and L.G. P.P and M.A.Z designed and built the fluidic, electrical, and optical setups. P.P designed and conducted the on-chip experiments and data analysis. M.A.J performed chemical sample preparation and column synthesis experiments. P.P. performed the numerical simulations. M.A.Z wrote the image processing codes to analyze bead fluorescence. Y.H. performed the contact angle and interfacial tension measurements. P.P. wrote the manuscript with significant inputs from M.A.Z, M.A.J and L.H.

**Competing Interests:** The authors declare no competing interests.

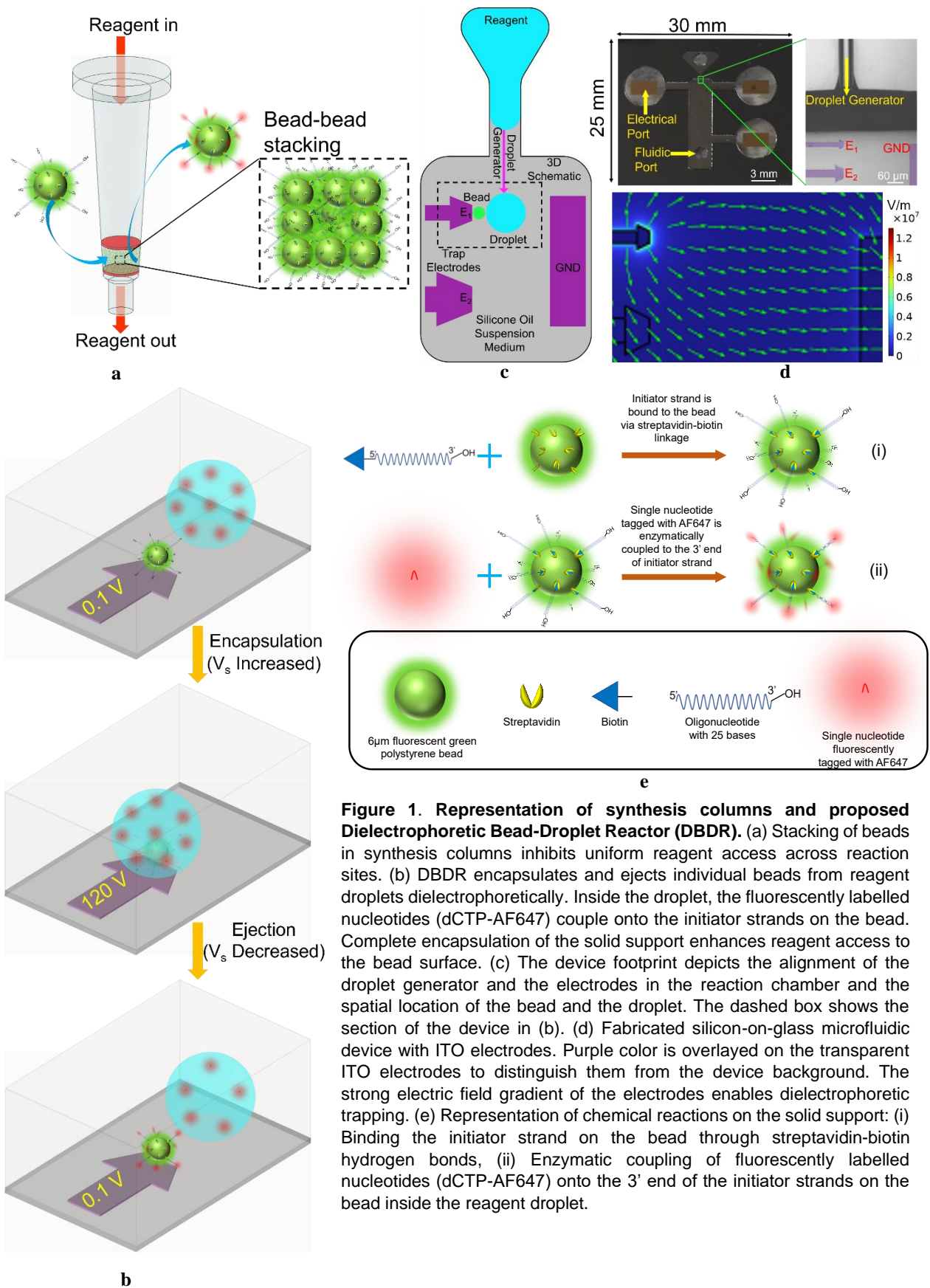

**Figure 1**. **Representation of synthesis columns and proposed Dielectrophoretic Bead-Droplet Reactor (DBDR).** (a) Stacking of beads in synthesis columns inhibits uniform reagent access across reaction sites. (b) DBDR encapsulates and ejects individual beads from reagent droplets dielectrophoretically. Inside the droplet, the fluorescently labelled nucleotides (dCTP-AF647) couple onto the initiator strands on the bead. Complete encapsulation of the solid support enhances reagent access to the bead surface. (c) The device footprint depicts the alignment of the droplet generator and the electrodes in the reaction chamber and the spatial location of the bead and the droplet. The dashed box shows the section of the device in (b). (d) Fabricated silicon-on-glass microfluidic device with ITO electrodes. Purple color is overlayed on the transparent ITO electrodes to distinguish them from the device background. The strong electric field gradient of the electrodes enables dielectrophoretic trapping. (e) Representation of chemical reactions on the solid support: (i) Binding the initiator strand on the bead through streptavidin-biotin hydrogen bonds, (ii) Enzymatic coupling of fluorescently labelled nucleotides (dCTP-AF647) onto the 3' end of the initiator strands on the bead inside the reagent droplet.

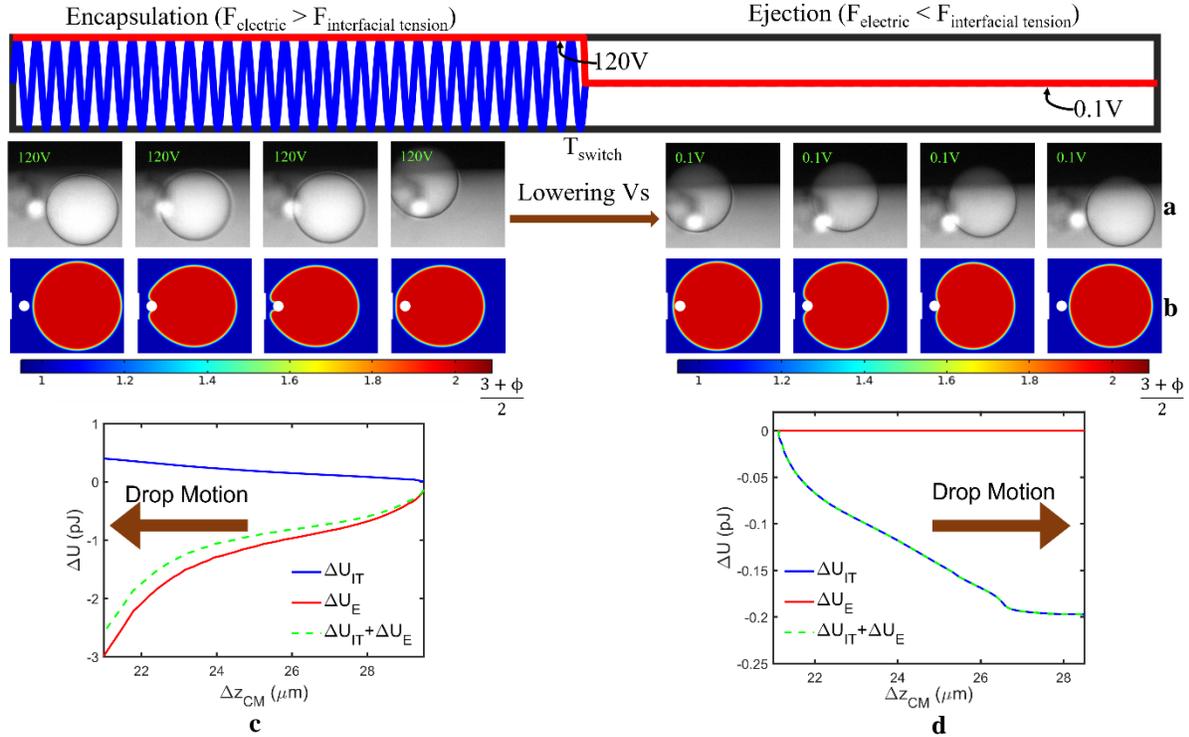

**Figure 2. Experimental demonstration and numerical simulation of the encapsulation and ejection of a microbead from a picoliter droplet.** (a) Experimental implementation and (b) numerical phase field simulation depicts the encapsulation of the bead by the droplet under high supply voltage ($V_s = 120\ V$) and its ejection from the droplet under low supply voltage ($V_s = 0.1\ V$). The phase field variable (ϕ) has a value of $-1$ in the silicone oil suspension medium (phase 1) and a value of 1 in the reagent droplet (phase 2). ϕ transitions from $-1$ to 1 at the droplet medium interface. The bead is represented in white. $T_{switch}$ is the time transition time from high voltage to low. (c) Electrocapillary potential energy representation of the engulfing and (d) ejection process. The droplet moves to minimize the total potential energy ($U = U_E + U_{IT}$) of the system. At high $V_s$, the change in electric energy stored between the electrodes is much larger than the change in Gibbs free energy due to the interfacial tension between the bead, droplet, and the suspension medium ($|\Delta U_E| \gg |\Delta U_{IT}|$). Therefore, $\Delta U \approx \Delta U_E$ decreases as the droplet approaches the electrode and encapsulates the bead. On the other hand, at low $V_s$, change in electric energy is much less than the change in Gibbs free energy ($|\Delta U_E| \ll |\Delta U_{IT}|$). Therefore, $\Delta U \approx \Delta U_{IT}$ decreases as the droplet moves away from the electrode and ejects the bead.

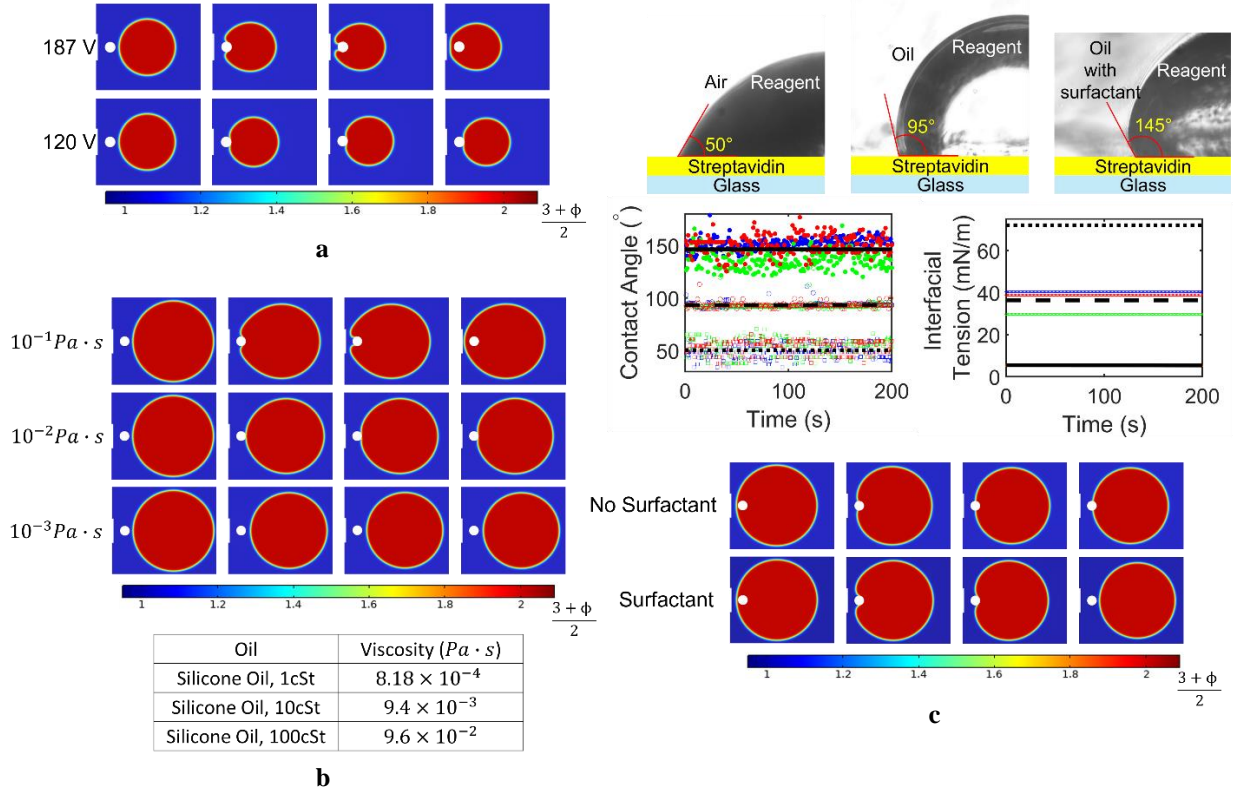

**Figure 3. System design.** (a) Smaller droplet ($R_d = 16\,\mu m$) requires a larger supply voltage ($V_s = 187\,V$) to overcome the capillary force and encapsulate the bead. At the lower voltage ($V_s = 120\,V$) the dielectrophoretic force cannot overcome the capillary force to encapsulate the bead into the smaller droplet (It was enough to encapsulate the bead in the larger droplet of $R_d = 25\,\mu m$, Figure 2). (b) The low viscosity of silicone oil 1cSt enables the ejection of the bead from the droplet at $V_s = 0.1$ V. With increase in oil viscosity the increased dissipation of the kinetic energy of the droplet prevents complete separation from the bead. (c) Silicone oil 1cSt (dashed line) renders the hydrophilic (in air, dotted line) streptavidin surface slightly hydrophobic. The surfactant Span80 further reduces the interfacial tension between the silicone oil and the reagent and increases the contact angle (solid line) that the reagent droplet forms on a streptavidin surface making it extremely hydrophobic It is critical for the ejection of the bead from the droplet. Interfacial tension and contact angle measurements are recorded with time variations to account for the surface adsorption. (Refer to Supplementary Material for details.) The streptavidin and glass thicknesses are not drawn to scale.

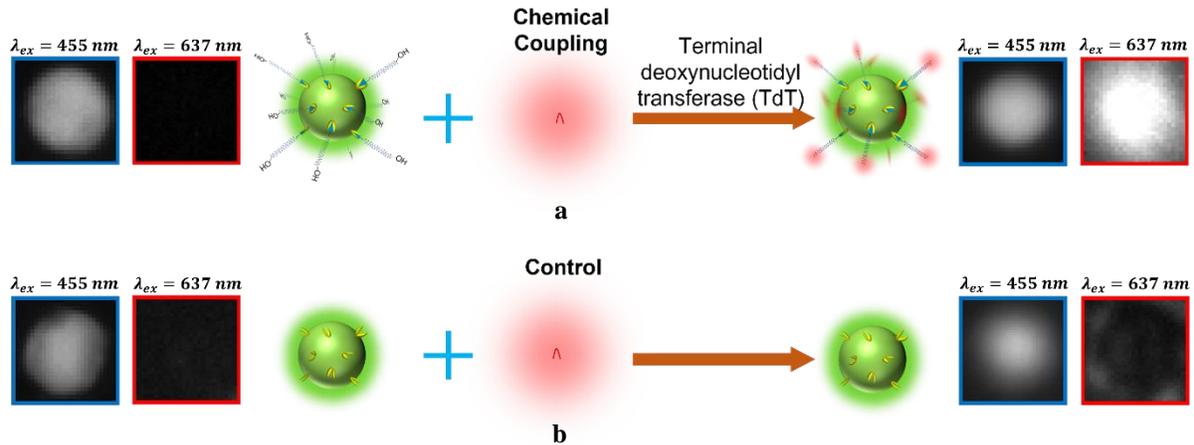

**Figure 4. Demonstration of enzymatic coupling of single nucleotides onto initiator strands on the bead.** (a) Enzymatic coupling of fluorescently labelled nucleotides (dCTP-AF647) onto the initiator strands on the streptavidin coated fluorescent green microbead. The nucleotides and the enzyme (TdT) are present in the reagent droplet. The bead has no red fluorescence prior to encapsulation within the droplet. However, after encapsulation and ejection from the droplet it fluoresces red indicating binding of the nucleotides to its surface. (b) Control experiment performed without initiator strands on the bead surface and enzymes in the droplet to establish the absence of any non-specific binding of the fluorescently labelled nucleotide onto the bead surface in (a). This is verified by the absence of red fluorescence from the bead after its ejection from the droplet. This confirms that the observed red fluorescence from the bead in (a) can be attributed to the chemical coupling of the nucleotide to the initiator strands on the bead.

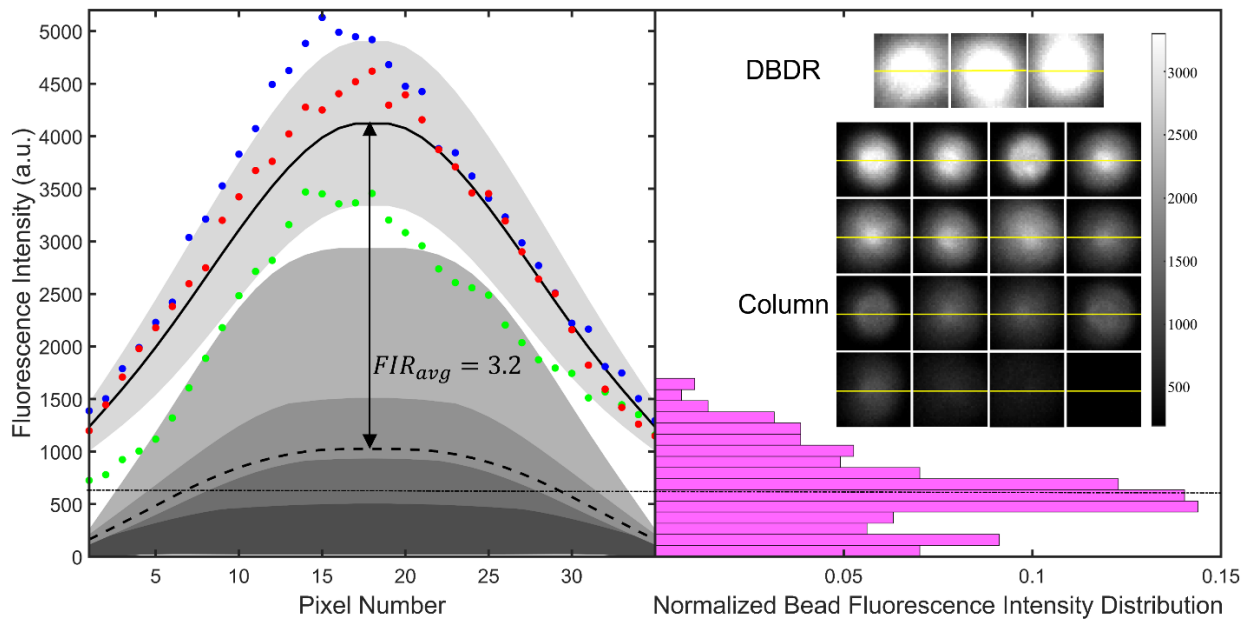

**Figure 5. Experimental Demonstration of Enhanced Enzymatic Coupling Fidelity of Nucleotide Using DBDR.** The lower grey band with brightness gradient in the left pane represents the fluorescence intensity distribution of beads reacted in the column. A few sample beads are shown in the inset on the right. The fluorescence intensities are taken along the yellow lines drawn across the beads. The brightness of the bands corresponds to the brightness of the beads in each row as well as to the quartiles of the histogram in the right pane which represents the distribution of the average fluorescence of all the beads reacted in the columns (a few sample frames of the fluorescence images of the beads reacted in columns are shown in Supplementary Figure 6a). The dashed curve in the left pane represents the fluorescence intensity distribution of the bead with average fluorescence intensity equal to the mean of the histogram. The mean ($\mu_{col}$) and the standard deviation ($\sigma_{col}$) of the average fluorescence intensities of beads reacted in the synthesis column are 629.27 and 373 respectively. These are evaluated over all the 282 beads. The three beads which represent the three experimental trials of DBDR are shown as insets in the right pane. Their mean ($\mu_{DBDR}$) and standard deviation ($\sigma_{DBDR}$) are 2057.5 and 209.05 respectively. Their fluorescence intensity distributions along the yellow lines are plotted as scatter plots on the left pane. The mean of the three distributions is represented by the solid black curve. The top grey band represents the standard deviation around the mean DBDR distribution.

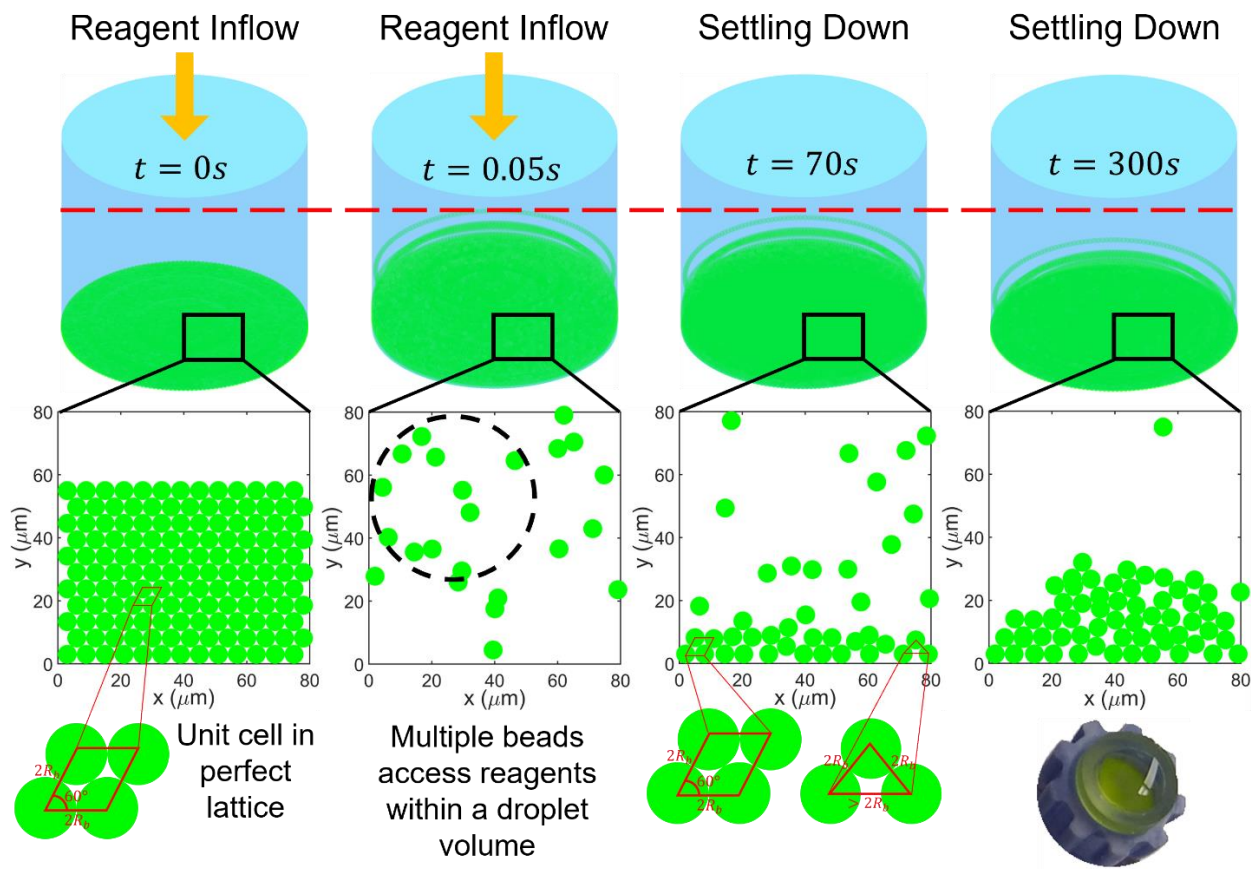

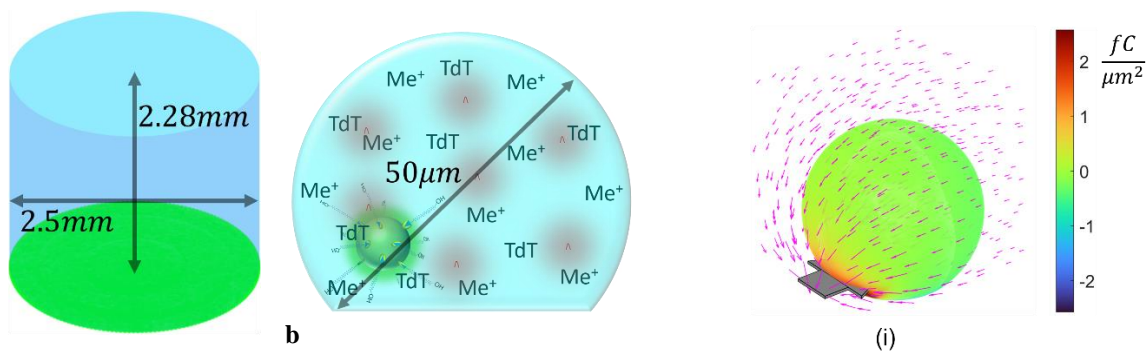

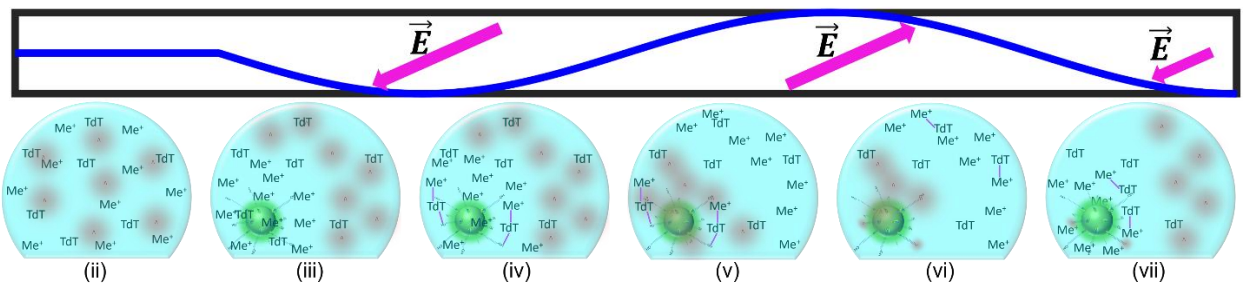

**Figure 6. Explaining enhanced fidelity of reactions in DBDR compared to columns.** (a) Stacking of beads prevents reagent access to reaction sites on the bead surface. Particle tracking simulations show that when the reagent is introduced into the column the beads are displaced from the stack due to turbulent fluid flow and then resettle over the experimental time once the reagent flow stops. This allows enhanced and varying reagent access

to many beads transiently as they settle down which explains the spread of fluorescence intensities of beads reacted in columns in Figure 5. The larger size of the column requires longer time ($T_{diff,col} \approx 25000s$) for the nucleotides (diffusion coefficient $D_{nucleotide} \approx 10^{-6}\ cm^2/s$) on the top to diffuse to the beads in the stack at the bottom. On the other hand, it takes much less time ($T_{diff,col} \approx 12.5s$) for the nucleotides to diffuse across the droplet. This can facilitate much faster reactions. (c) The applied AC electric field can also affect the migration of charge species involved in the reaction in the aqueous droplet and impact the reaction fidelity. (i) The applied non-uniform electric field due to the electrode leads to high density of induced surface charges on the droplet surface near the electrode (where the bead is trapped, Figure 1b & c and Figure 2a & b). (ii) In the absence of the field the charged reacting species are uniformly distributed within the droplet. $Me^+$ is a general representation of all positively charged metal cations which can be $Na^+$, $Mg^{2+}$, $Co^{2+}$ and $Zn^{2+}$. The fluorescent base is negatively charged due to its phosphate group. (iii) During the negative half of the AC supply cycle, the $Me^+$ions accumulate on the droplet surface closer to the electrode (where the bead is trapped) while the negatively charges nucleotides accumulate on the opposite side. (iv) The larger concentration of the ions facilitates their incorporation into the enzyme (TdT) and the subsequent binding to the initiator strand. (v) When the AC supply cycle switches, the negatively charged nucleotides drift towards the electrode while the unincorporated $Me^+$ and $Me^+ - TdT$ complexes that did not bind to the opposite end. (vi) The increased concentration of nucleotides in the vicinity of the beads facilitates faster reactions. (vii) As the AC supply switches again, the cycle continues.

# Supplementary Material for Bead-Droplet Reactor for High-Fidelity Solid-Phase Enzymatic DNA Synthesis

Punnag Padhy[1*], Mohammad Asif Zaman[1], Michael Anthony Jensen[2,3*], Yao-Te Cheng[1], Yogi Huang[4], Ludwig Galambos[1], Ronald Wayne Davis[2,3,5], Lambertus Hesselink[1*]

[1]Department of Electrical Engineering, Stanford University, Stanford, CA, 94305, U.S.A

[2]Stanford Genome Technology Center, Stanford University, Palo Alto, CA, 94304, U.S.A

[3]Department of Biochemistry, Stanford University, Stanford, CA, 94305, U.S.A

[4]Department of Chemical Engineering, Stanford University, Stanford, CA, 94305, U.S.A

[5]Department of Genetics, Stanford University, Stanford, CA, 94305, U.S.A

*Corresponding author. Email: punnag@stanford.edu, m.a.jensen@stanford.edu, hesselink@ee.stanford.edu

**This supplementary material document provides details of the device fabrication, experimental setup, sample preparation, experimental procedure, image and data analysis, and numerical and analytical calculations that complement the results discussed in the main manuscript.**

## 1. Device Fabrication

Fabrication started with ~ 500 μm thick silicon wafers (p-type, 10-20 Ω cm, <100>) and borofloat glass wafers ~550 μm thick that are 100 mm in diameter.

### 1.1. Silicon Component

The silicon wafer was etched in 5 layers to define the 1) alignment marks, 2) droplet generation microchannel, 3) reaction chamber 4) and 5) ports for external fluidic and electrical connections (Supplementary Figure 1a). The first four layers were defined by reactive ion etching using $SF_6$ gas. In the 4th layer the ports were etched almost through the wafer from the backside. In the 5th layer, the ports were completed by laser drilling through the ports from the front side. The photoresist for each layer was patterned using a standard photolithographic approach with a maskless exposure system (Heidelberg MLA 150).

### 1.2. Glass Component

800 nm of ITO (Indium Tin Oxide) was sputter deposited on piranha cleaned borofloat glass wafers at LGA Thin Films. This was followed by 2 layers of photlithoraphic processing: 1) the electrodes and alignment marks were defined by RIE of ITO from the entire wafer (barring the electrodes) using $CH_4$ and $H_2$ gases and 2) the contact pads were defined by evaporative deposition of 10 nm Cr and 200 nm Gold followed by metal liftoff. After each layer, the wafer was wet cleaned using $5:1:1::H_2O:H_2O_2:NH_4OH$ at 70°C for 1 hour (Supplementary Figure 1b).

### 1.3. Bonded Chip

The glass and silicon wafers were then aligned (Supplementary Figure 1c) and bonded anodically at 350°C by applying a voltage of 350V for 6 mins. The wafers are then diced into 25 mm x 30 mm chips using a laser cutter (Supplementary Figure 1d and e, and Figure 1b). The chips were silanized using vapor phase deposition of Dimethyldichlorosilane (DDMS) at Integrated Surface Technologies to make the device interior surface hydrophobic for the formation of water droplets. The silane on the exterior surface of the chip was stripped using UV ozone treatment to stick fluidic connectors using Loctite 401 adhesives (Supplementary Figure 1f). Electrical leads are soldered onto the gold contact pads through the electrical ports in the silicon wafer (Supplementary Figure 1f).

## 2. Experimental Setup

The experimental setup consists of the device holder interfaced with the fluidic, electrical, and optical subsystems.

### 2.1. Device Sample Holder

A 3D printed plastic sample holder was used to mount the device on the experimental setup. Copper pins hold the device in place with a 0.17 mm glass coverslip (24mm x 40 mm) from SPI below it. The sample holder was screwed onto a X-Y stage from Newport (Model#-406) mounted on a modified Nikon TE2000U inverted microscope (Supplementary Figure 2a and b).

### 2.2. Fluidic Subsystem

A piezo driven pressure controller (OB1 MK3+ from Elveflow) with a 30-psi input from the house nitrogen supply and a maximum output of 2000 mbar was connected to the input of a fluidic tank (15 ml plastic tube) using a 10 mm OD tubing (Supplementary Figure 2c and d). The output of the tank flows into the device via a 1/16-inch OD and 1/32-inch ID polytetrafluoroethylene (PTFE) tubing from Masterflex (item#-EW06407-41) which was plugged into the fluidic port of the device using the Nanoport Assemby from IDEX Health & Science (part# - N-333).

### 2.3. Electrical Subsystem

An A.C. high voltage amplifier (A. A. Lab Systems Ltd. Model#-A-303) amplifies the signal from a function generator (Hewlett Packard Model#-8116A) to generate a maximum output amplitude of 200 V (Supplementary Figure 2e and f). The amplified output was connected to the electrical leads of the device through a single pole double throw (SPDT) switch which connects the supply across either of the electrodes ($E_1$ or $E_2$) and the ground pad.

### 2.4. Optical Subsystem

The optical subsystem (Supplementary Figure 2g and h) was built around a modified Nikon inverted TE2000U microscope. A blue LED (SOLIS-445C from Thorlabs, 445 nm and 5.4 W min) with a band pass excitation filter (D480/30x from Chroma) images the bead (6μm diameter fluorescent green streptavidin coated polystyrene beads with excitation maxima at 441nm and emission maxima at 486nm, Catalog#-24157) and the fluid flow in the device onto a sCMOS camera from Thorlabs (Part#-CS2100M-USB). The experiments were recorded at 33 frames per second. A red LED (M625L4 from Thorlabs, 625 nm and 700 mW) excites the Alexa-647 to detect nucleotides labelled with the fluorophore (dCTP-AF647) in the reagent droplet. A sCMOS camera from PCO (PCO edge 5.5) was used to capture the low light intensity levels emanating from the nucleotides coupled to the initiator strands on the beads at 2s integration time. Appropriate bandpass excitation (Item#-86-988 from Edmund Optics, 640 nm center wavelength, 14 nm bandwidth, OD - 6) and emission (Item#-86-987 from Edmund Optics, 676 nm center wavelength, 29 nm bandwidth, optical density - 6) filters were used to ensure non-overlap of the excitation and emission spectrum. A Nikon objective (ELWD-20, 20x mag, 0.45 NA) with a correction collar for spherical aberration correction (set at 0.7mm which is 0.17mm thick glass coverslip + 0.53mm thick borofloat glass of the device) was used for imaging.

## 3. Sample Preparation

### 3.1. Preparing the oil solution by adding surfactant

4 ml of Span 80 (S6760 from Sigma Aldrich) was added to 200ml of 1 cSt silicone oil (PSF – 1cSt from Clearco Products) to make a 2.5% w/w solution. It was sonicated for 30 minutes to ensure complete dissolution of the surfactant.

### 3.2. Attaching initiator strand to beads

The initiator strand, which was a biotinylated oligomer with 25 bases (T25mer, 5' biotin, IDT) was attached to the 6 μm diameter ($R_b = 3\mu m$) streptavidin coated green-fluorescent polystyrene beads using the strong biotin-streptavidin hydrogen bond. The reaction was carried out for 60 min at 23°C, 14 RPM. The approximate starting yield (140 attomoles per bead (T25mer bound)) was determined by measuring the optical density at 260nm (Nanodrop) before and after initial binding, then subtracting the supernatant and wash OD values from the starting yield. Binding/ wash buffer: 20 mM Tris pH 7.5, 1 M NaCl, 1 mM EDTA, 0.0005% Triton-X 100 (45 μl (plus 5 μl 100 μM T25) for binding reaction, and 500 μl for wash steps). Based on a particle concentration of 1.4%, the number of beads was approximately 867,000 per 25 μl reaction (accounting for a 20% loss due to mixing and wash steps). Beads with initiator strands were then spun down using an Eppendorf Minispin (Catalog#-022620100) to remove the supernatant and were segregated into two parts (i) for Dielectrophoretic Bead-Droplet Reaction in the fabricated chip, and (ii) for benchtop synthesis in columns.

### 3.3. Suspending initiated beads in oil solution

The spun down initiated beads are suspended in the oil solution by sonication. The concentration of the beads in the silicone oil solution were tuned to ensure mostly a single bead floats in the vicinity of the electrodes within the field of view of the objective.

### 3.4. Preparing Reagent Solution

The reagent solution was prepared by mixing 25 µl of reagents consisting of the fluorescently labelled base (dCTP-AF647) in a buffer solution of 50mM Potassium Acetate, 20mM Tris-acetate, 10mM Magnesium Acetate, and 0.25mM Cobalt Chloride with the enzyme (TdT) solution consisting of 3 µl of 50mM KPO4, 100mM Sodium Chloride, 1.43 mM β-ME, 50% glycerol, and 0.1% Triton X-100 solution in an Eppendorf tube. This reagent solution was formulated by initial benchtop experiments as described in the subsequent experimental procedure section. A trace amount of sodium salt of fluorescein (F6377 from Sigma Aldrich) was added to the reagent using a toothpick to discriminate it from the continuous phase inside the microfluidic device.

### 3.5. Filling device with oil solution as continuous phase

The device was completely immersed in 50 ml of the 2.5% w/w solution of Span80 in 1cSt silicone oil contained in a glass jar inside a vacuum desiccator. As the desiccator was evacuated the air inside the device was drawn out. When the desiccator is refilled with air, the silicone oil solution gushes into the device to fill it completely without any trapped air bubbles.

### 3.6. Mounting device on sample stage

The device was then removed from the glass jar, its outer surface was cleaned by thoroughly wiping with isopropanol and then mounted on the sample holder (Supplementary Figure 3a). The objective was focused on the output of the droplet generation channel and the ITO electrodes.

### 3.7. Making electrical and fluidic connections to device ports

Electrical connections are made from the output of the amplifier to the ground pad and to the trap electrodes through the SPDT switch (Supplementary Figure 3a). The fluidic tank was filled with 15 ml of the above oil solution. A pressure was applied using the pressure controller to fill the output PTFE tubing from the tanker with oil solution which is dipped at the other end inside the 1.5 ml tube containing the reagent. Just before oil starts dripping from the tubing into the reagent tube, the height of the PTFE tubing and reagent tube were raised to suck the reagent into the tubing. Then the tubing was lowered again into another Eppendorf tube containing the oil solution. As the reagent solution started dripping, the height was raised again to fill the PTFE tubing with the oil solution while ensuring there are no trapped air bubbles. The tubing was then connected to the device inlet while pushing out the oil solution at the bottom to ensure no air gaps and fluidic continuity (the oil solution inside the device and at the bottom of the tubing are the same). This approach prevented immediate flow of the reagent solution through the droplet generation channel as soon as the PTFE tubing was connected thus allowing time for experimental setup and control (Supplementary Figure 3b).

## 4. Experimental Procedure

### 4.1. Encapsulation and ejection of bead from droplet

Beads suspended in the oil solution were introduced into the device through the oil inlet (Extended Data Fig. 1a). The voltage supply was switched on and set to around 15V amplitude at 25kHz to dielectrophoretically trap a bead floating near the top electrode (Supplementary Figure 3c). Then the voltage supply was switched off. Following this a pressure of 60 mbar was applied on the pressure controller to drive reagent flow in the device. As the reagent approached the entrance of the microfluidic channel, a sudden pressure pulse of 10 mbar was exerted to dispense a single droplet into the reaction chamber (Supplementary Figure 3d). Then the voltage supply was switched on again at15V amplitude and 200 Hz to trap the droplet adjacent to the bead on the top electrode (Supplementary Figure 3e).

At this point the supply voltage was gradually increased to ~ 120V amplitude. The droplet moves toward the electrode to encapsulate the bead. Subsequently the voltage was reduced to 0.1V amplitude and the bead was ejected out of the droplet. Supplementary Movie 1 depicts the encapsulation and ejection process (Figure 2).

### 4.2. Chemical coupling of base and control on the device

The above physical process was used (with a 300s encapsulation time) for the enzymatic coupling of fluorescently tagged nucleotides onto the initiator strand on the bead with a few additional intermediate images of the area around the reaction zone captured as enlisted below.

- Before loading the beads into the device an image to estimate the background noise under red illumination (Supplementary Figure 4a and b).
- After the bead was dielectrophoretically trapped on the top electrode an image each with the blue and red excitation are captured to measure the level of the red fluorescence signal from the site of the bead just prior to the reaction (Supplementary Figure 4c and d).
- Finally, after the encapsulation and ejection process another set of images under blue and red excitation were captured (Supplementary Figure 4e and f).

These images were captured as 16-bit Tiff files. Supplementary Figure 4c-f are used in Figure 4a of the main text. These steps were repeated to see the repeatability of the chemical coupling reaction on our platform. The three different reacted beads under red excitation are depicted separately in Figure 4g and are also used in Figure 5 of the main text.

The control experiment was repeated using the above process but with beads without initiator strands and reagent droplets without the enzymes (Figure 4h-m). Supplementary Figure 4j-m are used in Figure 4b of the main text.

## 4.3. Column Synthesis

Firstly, free solution reactions are implemented to develop optimal room temperature protocol (Figure 5a) for translation into DBDR. Results were analyzed using reverse-phase high performance liquid chromatography (HPLC) (Figure 5b).

To simulate an enzymatic synthesis reaction using a column, an open-top nylon syringe filter (Omicron SFNY04XB, 4 mm, 0.45 µm) was used. To the bottom filter (0.45 µm pore size) which was held in place by a plastic ring, 15ul of beads (1.2 M) were added. A top filter was then positioned above the reagent bed. Between the filters the reaction volume was about 15 µL. A 1ml syringe was used to push the bead medium (10mM HCL, 2M NaCl, 1mM EDTA, 0.0005% Triton-X 100, pH 7.3) passed the bottom filter until it completely exited the drip director. Fifty µL reagent (6 µL TdT (20U/ul), 5 µL 10x TdT buffer (50 mM potassium acetate, 20 mM Tris-acetate, 10 mM magnesium acetate, pH 7.9 @ 25°C), 5 µL 10X (2.5 mM) solution of $CoCl_2$, 0.25 µL 1 mM Alexa Fluor™ 647-aha-dCTP, 33.75 µL water) were added to the top filter, and an empty 1ml syringe was used to push the reagent passed the top filter (flow rate at 50 µL per second), into the reaction area until the reagent could be seen inside the drip director. The column was kept upright for 5 minutes. Once the reaction was completed, the empty 1ml syringe was used to push the spent reagents through the column until the drip director was clear. Afterwards, 1 ml of bead suspension medium (described above) was used to wash the beads. These beads were than taken in a 1.5 ml Eppendorf tube for analysis using fluoroscopy.

As a control, synthesis was performed on beads without initiators and reagents without TdT and the reaction was analyzed through fluorescence measurements.

## 4.4. Measuring fluorescence from beads reacted in columns

About 3µl of the reacted bead suspension in the buffer was taken in an Eppendorf tube and was diluted to ensure the bead concentration was small enough to prevent signal interference from beads in different planes while being large enough to have ample beads within the field of view to get a statistically significant inference about fluorescence intensity distribution. The beads were introduced into the device filled with MilliQ water. The same chips were used for the fluorescence measurements to ensure identical optical environment for comparison between on-chip experiments with their column counterparts. Once the beads settled down (imaged using blue excitation) the excitation was switched to red to image the fluorescence intensity of the beads. Many such frames of red fluorescent beads were collected with a large number of beads ($\approx$ 300). A select few frames are shown in Supplementary Figure 6a. A few representative beads spanning the entire range of fluorescence intensities are used in Figure 5 of the main text.

The control experiments implemented using DBDR (Figure 4b of main text) were reiterated on the columns. The fluorescence of these beads was measured following the same procedure as discussed in the previous paragraph (Supplementary Figure 6b).

### 4.5. Estimating interfacial tension and contact angle

The interfacial tension between the reagent droplet and the silicone oil solution (with 2.5% w/w Span80) reported in Figure 3 of the main text were measured using the standard Wilhelmy plate method. The contact angle that the reagent droplet forms on a streptavidin coated surface (the polystyrene bead used in the experiments is coated with streptavidin) in the above silicone oil suspension medium was measured by capturing the droplet shape on a streptavidin coated glass slide (GS-SV-5 from Nanocs) and estimating the angle it forms on the surface of the slide through shape fitting.

The contact angle that the reagent droplet forms on the inner walls of the silanized device is estimated using optical microscopy. After generating the reagent droplet, the objective is focused on the electrode and the radius of the droplet at the glass surface is measured. After that the focus is shifted up till the sharpest image of the droplet boundary is formed. The radius of the greater circle of the droplet is measured. Using these two measurements and simple trigonometry the contact angle is estimated to be 137.9° (Supplementary Figure 7). As a confirmation of this again using elementary trigonometry the height of the greater circle from the glass surface is estimated to be $18.3 \mu m$ and closely matches the shift in the focus of the objective.

### 4.6. Estimating electrical conductivity of reagent

The conductivity of the reagent droplet was measured using an Orion 3 Star Conductivity Portable.

## 5. Image Processing and Data Analysis

### 5.1. Encapsulation and ejection of bead from droplet

The recorded video (Movie 1) of the encapsulation and ejection process was analyzed frame by frame using ImageJ and snapshots that best represent the processes were selected and labelled for Figure 2 of the main text.

### 5.2. Establishing enzymatic coupling of base to the initiator strands on the bead in DBDR

Maintaining the same scale of 250-3300 across the red fluorescence images, the difference in brightness of the bead with enzymatic coupling (Figure 4a main text and Supplementary Figure 4f) and the control bead (Figure 4b main text and Supplementary Figure 4m) was obvious.

### 5.3. Analyzing fluorescence intensity distribution

Each frame in Supplementary Figure 6a (one such frame shown in Supplementary Figure 8a) was analyzed using predefined image processing functions in Matlab to detect the beads, binarize them (one such frame shown in Supplementary Figure 8b), evaluate their mean fluorescence intensity, and evaluate fluorescence intensity distribution across a horizontal line passing through the bead center (one such frame shown in Supplementary Figure 8c). Average fluorescence values across frames were collected to plot a histogram of the fluorescence intensity distribution of all beads reacted using synthesis columns (Supplementary Figure 8d).

## 5.4. Analyzing statistical distribution of data

The average fluorescence intensity of the three beads reacted using DBDR were found to be 2184.5, 2171.7, 1816.2 (same unit as was used for column fluorescence data). To validate the statistical significance of our results, we performed the Welch's t-test on both the data sets using the inbuilt ttest2 function in Matlab under the null hypothesis that the distributions have identical means. The null hypothesis was rejected at 0.05 significance level ($P = 1.99 \times 10^{-10}$). Therefore, the means of the distributions are non-identical. Furthermore, the statistical power of the test was 0.9997 (evaluated using inbuilt Matlab functions sampsizepwr) for confidence interval of 98%. So, the sample size is enough to reject the null hypothesis at 0.02 significance level with almost certainty.

# 6. Numerical and Analytical Modelling

## 6.1. Electrohydrodynamic Simulation of the Encapsulation and Ejection of the Bead by the Droplet

We modeled the electric field-driven engulfing of the bead by the droplet and its subsequent ejection by coupling the Navier Stokes equation for incompressible fluids [13] (eq. S1) with the charge continuity equation of electrodynamics [13] (eq. S2). The spatially varying electric field due to suitably defined electrodes exerts non-uniform pressure on the fluidic interface resulting in a net force on the aqueous reagent droplet suspended in a medium of silicone oil [13,19] with a kinematic viscosity of 1cSt. The electric force exerted on the fluidic interface is given by the Maxwell Stress Tensor [19] (eq. S3). This electric force drives fluid flow as per the Navier Stokes equation [13] (eq. S1). As a result, the fluidic interface shifts with time. The shift in fluidic interface with time is tracked by the phase-field method [S1] (eq. S4). This shifting fluidic interface is fed back into the electrical charge continuity equation (eq. S2) as a change in the material boundary and hence a change in the electric boundary condition. Mimicking experimental observations (Figure 2(a) in the main text) the bead was kept stationary in the simulations. This also reduces the number of moving components and greatly simplifies the simulation while keeping the essential physics underlying the process intact. Furthermore, we employ an axis-symmetric model for our simulation to reduce computational resource requirements (Supplementary Figure 9a) while focusing on the essential physical principles underlying the process.

$$\rho \frac{\partial \vec{u}}{\partial t} + \rho(\vec{u} \cdot \vec{\nabla})\vec{u} = \vec{\nabla} \cdot [-p\overleftrightarrow{I} + \overleftrightarrow{K}] + \vec{F}_{electric} \qquad \text{(eq. S1a)}$$

$$\rho \vec{\nabla} \cdot \vec{u} = 0 \qquad \text{(eq. S1b)}$$

$$\overleftrightarrow{K} = \mu \left\{ (\vec{\nabla}\vec{u}) + (\vec{\nabla}\vec{u})^T \right\} \qquad \text{(eq. S1c)}$$

$$\vec{F}_{electric} = \vec{\nabla} \cdot \overleftrightarrow{T}_{electric} \qquad \text{(eq. S1d)}$$

Here ρ is density of the fluid, $\vec{u}$ is velocity of fluid flow, p is pressure, $\overleftrightarrow{I}$ is the identity tensor, $\overleftrightarrow{K}$ is the viscous stress tensor, μ is the dynamic viscosity of fluid and $\overleftrightarrow{T}_{electric}$ is the electric component of Maxwell Stress Tensor.

$$\vec{\nabla} \cdot \vec{J} = -\frac{\partial \rho_q}{\partial t} \tag{eq. S2a}$$

$$\vec{J} = \sigma \vec{E} \tag{eq. S2b}$$

$$\vec{E} = -\vec{\nabla} V \tag{eq. S2c}$$

$$\sigma = Vf_r\sigma_r + Vf_o\sigma_o \quad \text{and} \quad \varepsilon = Vf_r\varepsilon_r + Vf_o\varepsilon_o \tag{eq. S2d}$$

Here $\vec{J}$ is the current density, $\rho_q$ is the charge density, σ is the electrical conductivity, $\vec{E}$ is the electric field, $\vec{D}$ is the electric displacement, $\varepsilon_0$ is the permittivity of free space, ε is the relative permittivity, $V$ is the electric potential, $Vf_{r/o}$ represents the volume fraction of the reagent droplet/oil medium, $\sigma_{r/o}$ and $\varepsilon_{r/o}$ represent the electrical conductivity and the electrical permittivity of the reagent droplet/oil medium respectively.

$$\overleftrightarrow{T} = \vec{E} \otimes \vec{D} - \frac{1}{2}(\vec{E} \cdot \vec{D}) \tag{eq. S3}$$

Here ⊗ represents outer product of two vectors.

$$\frac{\partial \phi}{\partial t} + \vec{u} \cdot \vec{\nabla}\phi = \vec{\nabla} \cdot \frac{3\chi\epsilon_{pf}}{2\sqrt{2}} \vec{\nabla}\psi \tag{eq. S4a}$$

$$\psi = -\vec{\nabla} \cdot \epsilon_{pf}^2 \vec{\nabla}\phi + (\phi^2 - 1) \tag{eq. S4b}$$

$$\vec{F}_{st} = \frac{3\gamma}{2\sqrt{2}\epsilon_{pf}} \psi \vec{\nabla}\phi \tag{eq. S4c}$$

$$Vf_o = \frac{1-\phi}{2} \quad \text{and} \quad Vf_r = \frac{1+\phi}{2} \tag{eq. S4d}$$

Here ϕ represents the phase field variable which is -1 in the suspension medium and 1 in the reagent droplet and transitions from -1 to 1 at the droplet-medium interface. The phase field method models the interface as a transition region of non-zero thickness over which the two fluids mix with varying volume fractions $Vf_r$ and $Vf_o$. $\epsilon_{pf}$ is the interfacial thickness parameter which determines the stiffness of the transition from the reagent phase (ϕ = 1) to the oil phase (ϕ = −1). It should be small enough to maintain the sharpness of the interface and capture the physics accurately but large enough to prevent the finite mesh size from causing numerical instabilities. χ is called the mobility tuning parameter. It should be large enough to accurately track the shift in the interface position with fluid flow while still being small enough to give a sharp enough of an interface. $\gamma$ is the surface tension coefficient.

As the droplet moves under the influence of the applied electric stress (eq. S1d and eq. S3) as per the Navier Stokes equation (eq. S1a-c) the spatial position of the droplet-medium interface evolves. The resultant evolution of the fluidic interface was tracked using the phase field method (eq. S4). The changing boundary was reflected as a change in the spatial profile of the electrical conductivity and relative permittivity (eq. S2d) which in turn modified the solution of the charge continuity equation (eq. S2a). This again results in a change in the force acting on the droplet (eq. S1d and eq. S3). This system of coupled equations was solved using COMSOL Multiphysics which is a commercially available finite element method-based simulation package. The user defined material properties used in the simulation are summarized in Supplementary Table 1.

## 6.2. Potential energy-based evaluation of the engulfing and ejection process

The Gibbs free energy of the fluidic interfaces is given by [13, 25, 26]:

$$U_{IT} = \gamma_{ow}A_{ow} + \gamma_{os}A_{os} + \gamma_{ws}A_{ws} \qquad \text{(eq. S5)}$$

The total surface area of the bead is a constant ($A_s$) (Supplementary Figure 9b). So,

$$A_s = A_{os} + A_{ws} \qquad \text{(eq. S6)}$$

The contact angle that a reagent droplet forms on the surface of the streptavidin coated bead in a medium of silicone oil is given by [13, 25, 26]:

$$\cos\theta = \frac{\gamma_{os} - \gamma_{ws}}{\gamma_{ow}} \qquad \text{(eq. S7)}$$

Using eq. S6 and eq. S7 in eq. S5 we obtain:

$$U_{IT} = \gamma_{os}A_s + \gamma_{ow}(A_{ow} - \cos\theta\, A_{ws}) \qquad \text{(eq. S8)}$$

The first part of the right side of eq. S8 is a constant. Therefore,

$$\Delta U_{IT} = \gamma_{ow}(\Delta A_{ow} - \cos\theta\, \Delta A_{ws}) \qquad \text{(eq. S9)}$$

Specifically for the system under consideration with $r_b = 3\mu m$ (radius of bead) and $r_d = 25\mu m$ (radius of droplet), as the droplet completely engulfs the bead starting from separate bead and droplet we have:

$$\Delta A_{ws} = 4\pi \times 3^2 \mu m^2 \qquad \text{(eq. S10)}$$

As the droplet engulfs the bead its volume remains constant and therefore its surface area with the surrounding medium increases. This can be expressed as follows:

$$\frac{4}{3}\pi(R_d + \Delta R_d)^3 = \frac{4}{3}\pi R_d^3 + \frac{4}{3}\pi r_b^3 \quad \text{(eq. S11a)}$$

$$\Delta A_{ow} = 4\pi(R_d + \Delta R_d)^2 - 4\pi R_d^2 \quad \text{(eq. S11b)}$$

From eq. S11A $\Delta R_d = 0.01 \mu m$. Therefore, $\Delta A_{ow} = 4\pi \times 50.0144 \times 0.0144 \mu m^2$. Comparing the terms on the right side of eq. S9 we see:

$$\frac{-\cos\theta \Delta A_{ws}}{\Delta A_{ow}} = 10.82 \quad \text{(eq. S12)}$$

Therefore, we neglect $\Delta A_{ow}$ in our calculations. So, we assume:

$$\Delta U_{IT} \approx -\gamma_{ow} \cos\theta \, \Delta A_{ws} \quad \text{(eq. S13)}$$

As the droplet moves the electrical energy of the system changes. That means the electrical energy stored in the circuit changes[S2].

$$\Delta U_E = -\frac{\Delta(QV_s)}{2} \quad \text{(eq. S14)}$$

Here $Q$ is the charge on the electrode and $V_s$ is the supply voltage on it. The total change in the energy of the system as the droplet encapsulates and ejects the bead is given by:

$$\Delta U = \Delta U_{IT} + \Delta U_E \quad \text{(eq. S15)}$$

As the droplet moves its center of mass shifts which was evaluated using the following equation for all $z$ such that $\phi > 0$ in the simulation.

$$z_d = \frac{\sum \rho z \Delta V}{\sum \rho \Delta V} \quad \text{(eq. S16)}$$

The center of mass of the bead is fixed ($z_b$) in our simulations. The distance between the center of masses is then evaluated as:

$$\Delta z_{CM} = z_d - z_b \quad \text{(eq. S17)}$$

The change in energies is plotted against $\Delta z_{CM}$ in Figure 2c and d of the main text.

### 6.3. Particle tracking simulations

These simulations were carried out to explain that the ideal case of perfect bead-bead stacking (which represents the case of minimum exposure of solid support bound growing strands of oligonucleotides to the influx of synthesis reagents) is seldom achieved in synthesis columns. To this end, we implement particle tracking simulations to show the displacement of beads from perfectly stacked initial positions due to turbulent motion of the reagent fluid as they are injected into the synthesis columns. As the beads move away from a perfectly stacked configuration, they have a higher access to reagents which decreases again as the beads eventually settle down after the reagent injection stops into a stacked configuration. So, overall, the reaction fidelity and the fluorescence intensity on the beads lies between the lower limit of a perfectly stacked configuration and the upper limit of maximum exposure as in the case of DBDR.

For simplicity of setup and reduced memory and simulation time requirements 2D mirror symmetric simulations were setup in COMSOL Multiphysics (Supplementary Figure 10). As a further simplification, instead of a completely coupled turbulent flow-particle tracking simulations, the steady state solution of the turbulent flow equations was obtained using the Reynolds-averaged Navier-Stokes (RANS) model for turbulence [S3]. This steady state solution of the turbulent flow equations was used to exert drag force [13] on the particles (eq. S18) in the time dependent particle tracking simulations. The particles also experienced the force of gravity (eq. S19). The fluctuating components of the turbulence velocities are represented using a continuous random walk model. Assuming that the solutions in the cylindrical columns would be azimuthally symmetric, the solutions of the particle tracking simulations were revolved around the symmetry axis (Supplementary Figure 10) to generate the plots in Figure 6a of main text.

Once the fluid flow subsides, the particles settle down under the force of gravity into a stack.

$$\vec{F}_{drag} = \frac{m_p}{\tau_p}(\vec{u}' - \vec{v}_p)$$ (eq. S18)

$$\vec{F}_{gravity} = \frac{\rho_p - \rho_f}{\rho_p} m_p \vec{g}$$ (eq. S19)

Here $m_p$ is the mass of a particle, $\rho_p = 1050 \ kg/m^3$ is its density, $\rho_f = 1000 \ kg/m^3$ is the density of the fluidic suspension medium, $\vec{g} = 9.8 \ kg/m^2$ is the acceleration due to gravity, $\tau_p = \frac{\rho_p d_p^2}{18\mu}$, $\vec{v}_p$ is the velocity of a particle and $\vec{u}' = \vec{u} + \vec{u}_f$ is the velocity of the fluid flow which is the sum of the mean flow velocity ($\vec{u}$) and the fluctuation term ($\vec{u}_f$).

## Supplementary References

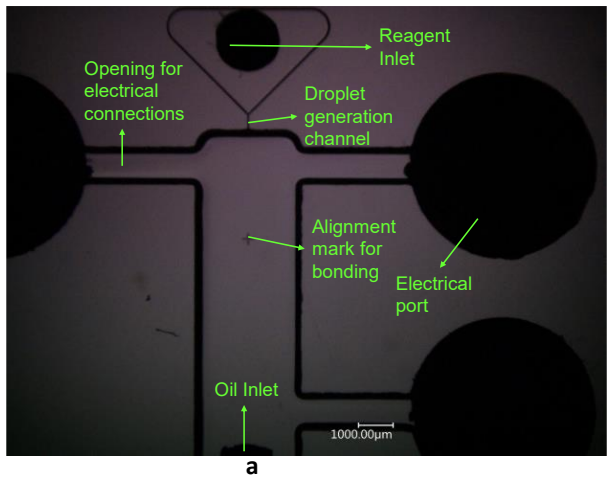
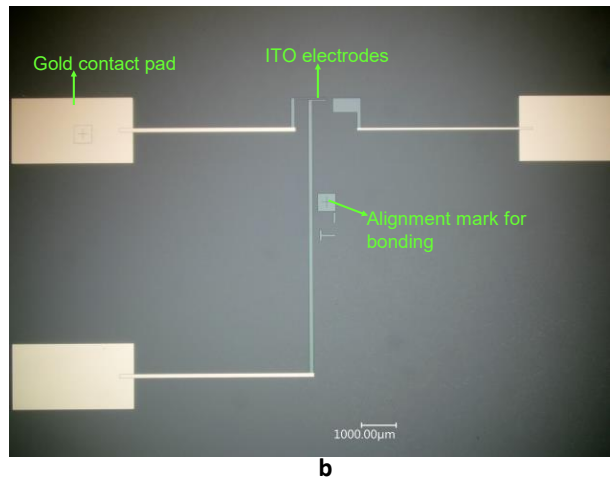
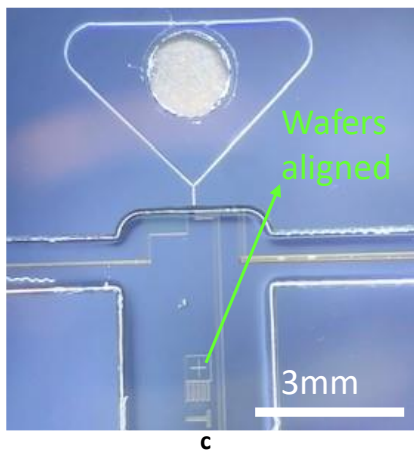
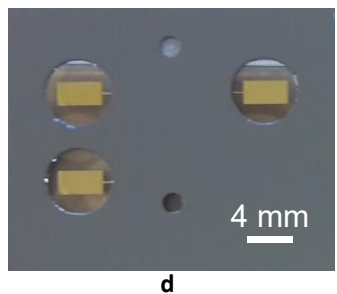
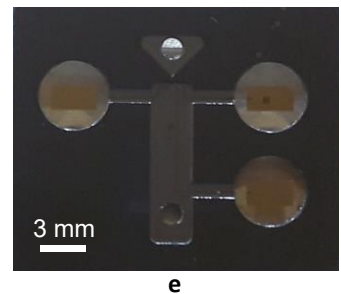
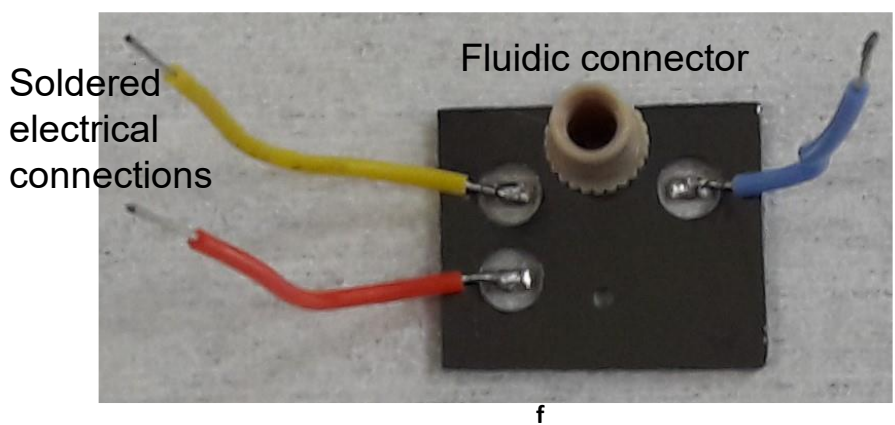

**Supplementary Figure 1. Device Fabrication. a,** Fabricated silicon and **b,** glass part. **c,** The silicon and the glass part of the device are aligned using the alignment markers in an EV Aligner system. They are then bonded and diced into chips. **d,** The top (port) view and the **e,** bottom (device) view of the fabricated silicon-on-glass microfluidic device. **f,** External electrical and fluidic connectors attached to the device ports.

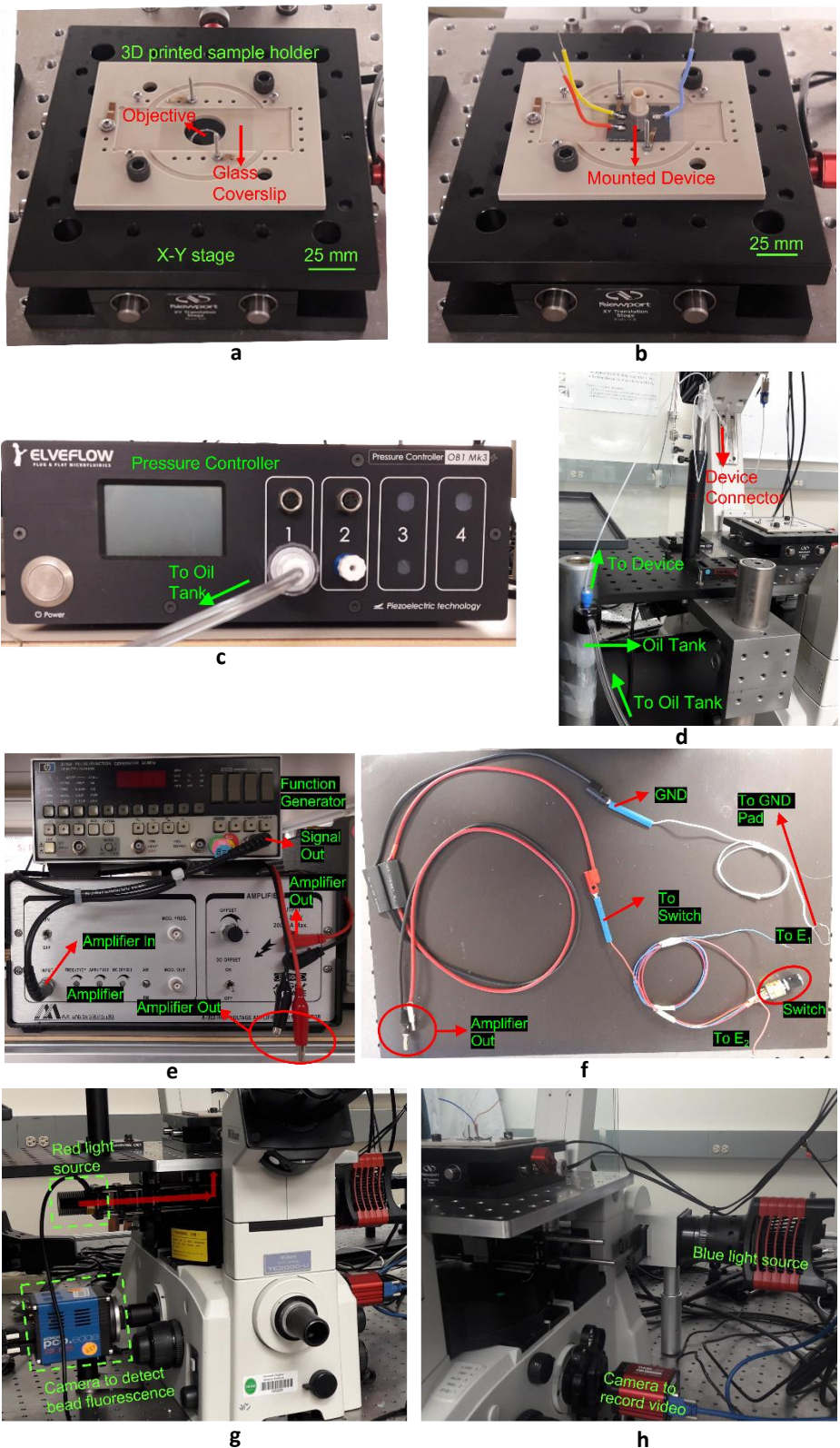

**Supplementary Figure 2. Experimental Setup a**, The 3-D printed sample holder is screwed onto the x-y stage which sits on an inverted microscope setup. **b**, The device is mounted on the sample holder on top of a glass coverslip to prevent any fluid flow onto the objective. **c**, The piezo-driven pressure controller (OB1 MK3+) from Elv-

Eflow drives fluid flow through the microfluidic device. The output of the pressure controller is connected to the oil tank. **d**, The pressurized air from the pressure controller drives fluid flow from the oil tank into the device. The device connector connects the output from the oil tank to the device. **e**, The electrical signal (upto 15V amplitude) from the function generator is fed to the amplifier which amplifies the signal by 20x. **f**, The amplified output of the signal generator is supplied to either of the electrodes ($E_1$ or $E_2$) using a single pole double throw switch. **g**, The red light source is used for the fluorescent detection of the labelled nucleotide. **h**, The blue light source is used for imaging fluid flow, the fluorescent green beads, and the encapsulation and ejection of the bead by the droplet.

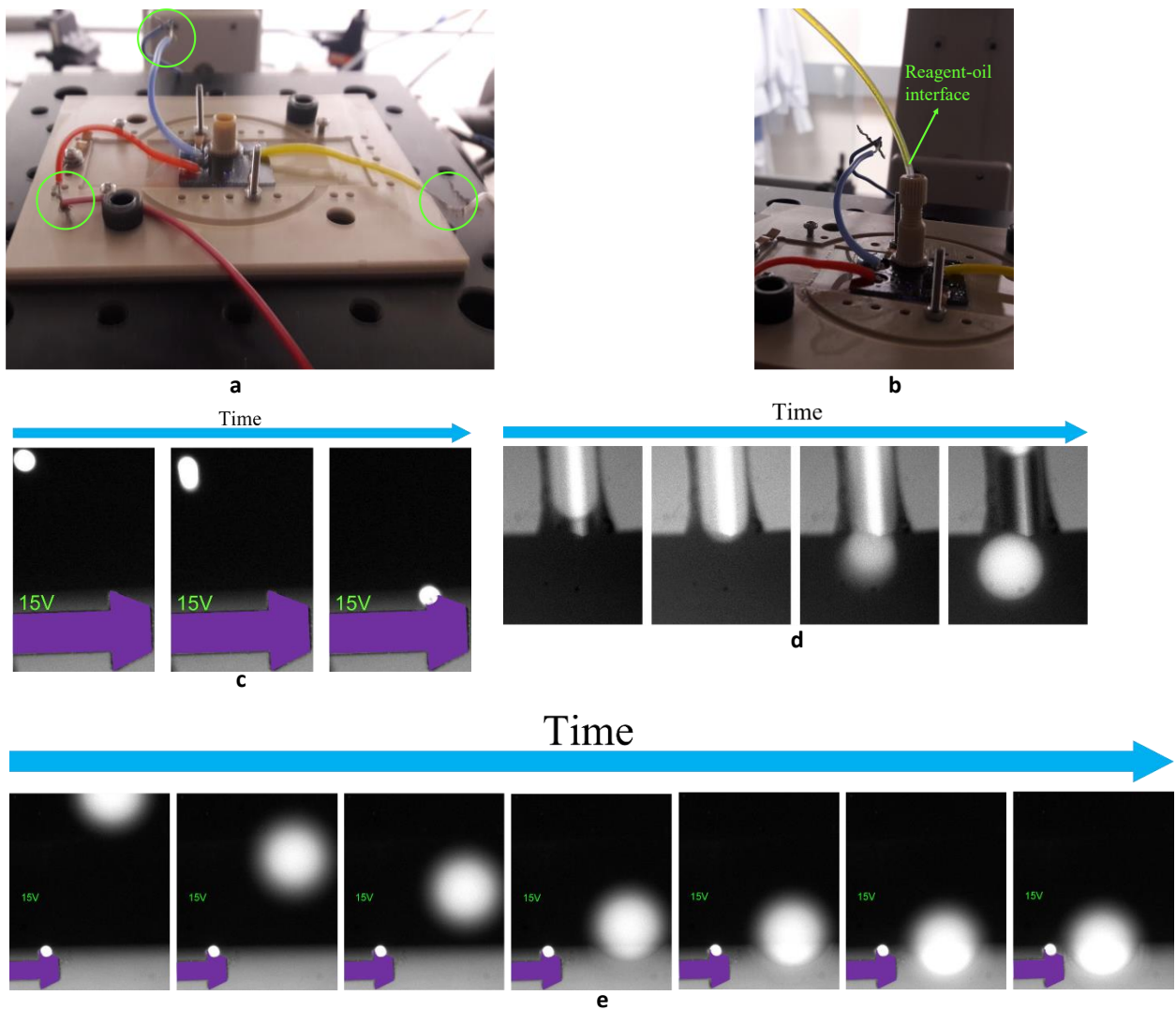

**Supplementary Figure 3. Sample Mounting, connecting external supplies, bead trapping, droplet generation and droplet trapping. a,** Electrical connections made to the three device terminals as shown within the three green circles. **b**, Fluidic connection made to the device using the N-333 connectors. The bottom of the tubing is filled with oil to ensure the reagent does not flow into the device as soon as the connection is established for better experimental control. The reagent flows in after the oil. The reagent-oil interface in the input tubing is shown. **c,** The bead is trapped dielectrophoretically by applying a voltage of $15\ V$ (amplitude) on the trap electrodes at a supply frequency of $25\ kHz$. The transparent ITO electrodes appear the same color as the background making it difficult to segregate in the image. Hence, false color has been overlayed (purple) to distinguish the electrode from the rest of the device background. **d**, Droplet generation in the presence of Span80 in the silicone oil suspension medium. Reduced interfacial tension between the reagent and the silicone oil suspension medium as well as the increased contact angle that the reagent forms with the device walls aids in the droplet breakup from the reagent fluid in the microchannel. **e**, The droplet is trapped dielectrophoretically by applying a voltage of $15\ V$ (amplitude) on the trap electrodes at a supply frequency of $200\ Hz$.

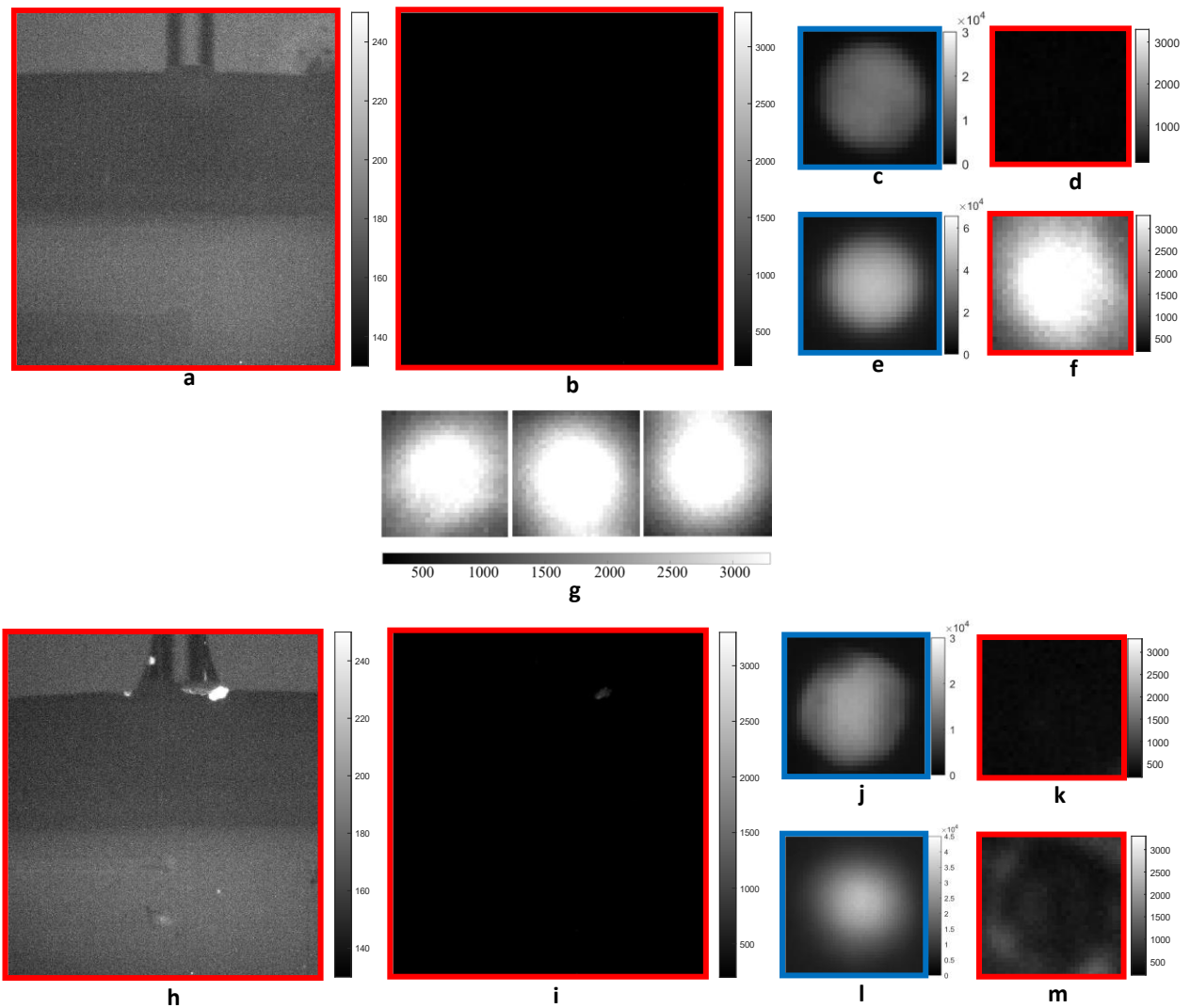

**Supplementary Figure 4. Fluorescence images and intensity data for enzymatic coupling and control experiment. a-f,** Coupling reaction. **g**, Red fluorescence images of beads from three consecutive reactions. **h-m,** Control reaction. **a** and **h**, Device filled with oil and with relevant electrical and fluidic connectors before start of experiment with color scale adjusted to lower thresholds and **b** and **i**, higher thresholds. **a** and **b** (**h** and **i**), are the same image with different color scales. The lower color scale in **a** and **h** is used to highlight the device structure under red illumination while the higher color scale in **b** and **i** is used to depict the absence of any source of noise at the level of the fluorescence signal from the reacted beads due to AF-647 used to label dCTP. **c** and **j**, Bead before encapsulation and ejection from droplet under blue and **d** and **k,** red illumination. **e** and **l**, Bead after encapsulation and ejection from droplet under blue and **f** and **m**, red illumination. The color of the boxes denotes the illumination source (blue $\lambda_{ex} - 455nm$, red $\lambda_{ex} - 637$nm).

| Reagent | Ml |
| --- | --- |
| TdT (20 U/ µl) | 3.0 |
| Buffer (10 x) | 2.5 |
| CoCl$_2$ (2.5 mM 10 x) | 2.5 |
| Beads (7 µl) or initiator (50 µM*) | 0 (or 1) |
| Dye (1 mM) | 0.1 |
| Water* | 16.9 (or 15.9) |
| **Total** | **25** |

a

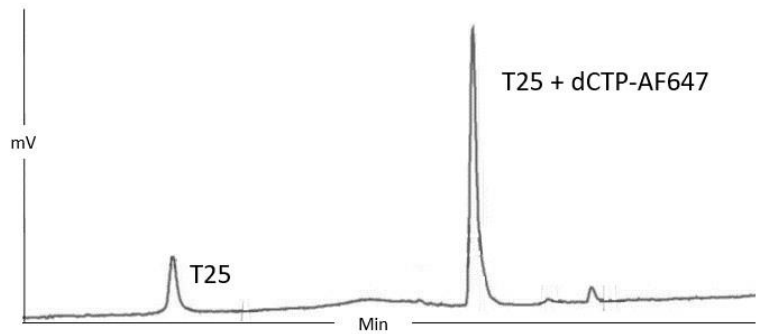

b

**Supplementary Figure 5. Benchtop synthesis reaction. a,** Either the reaction was prepared 1) for synthesis free in-solution (no beads) for parameter optimization using high performance liquid chromatography (HPLC) where 1 µl (50 µM strand (T25) was added to the reaction mix or 2) for synthesis directly onto the beads (in this case 7 µl beads with the initiator already attached were used); for experiments performed in a column (Fig. 5 and 6 of main text), reaction volumes were doubled to 50 µl. In both cases (1 and 2), 10 M EDTA was used to stop the synthesis reaction after 5 min. Buffer contents: 50 mM potassium acetate, 20 mM Tris-acetate and 10 mM magnesium acetate; **b**, HPLC chromatogram showing general results for post-synthesis on the benchtop (5 min, 23°C) free in-solution (T25mer coupled to dCTP-AF647); HPLC analysis was only performed to optimize synthesis conditions before translating to the device. Parameters for sample processing include, Hitachi WAV system, DNA-Sep column (C-18, Cat. DNA-99-3510); method conditions: Buffer A (0.1 M TEAA (triethylammonium acetate) in water), Buffer B (0.1 M TEAA (ADS Biotech), 25% acetonitrile (Sigma Aldrich)), 80°C, 1.2 ml/ min, gradient: 82% to 12% A for 10min; absorbance and fluorescence were measured at 260 nm and 648 nm (excitation)/ 688 nm (emission), respectfully. Five µl directly aspirated from the reaction were added to the HPLC column.

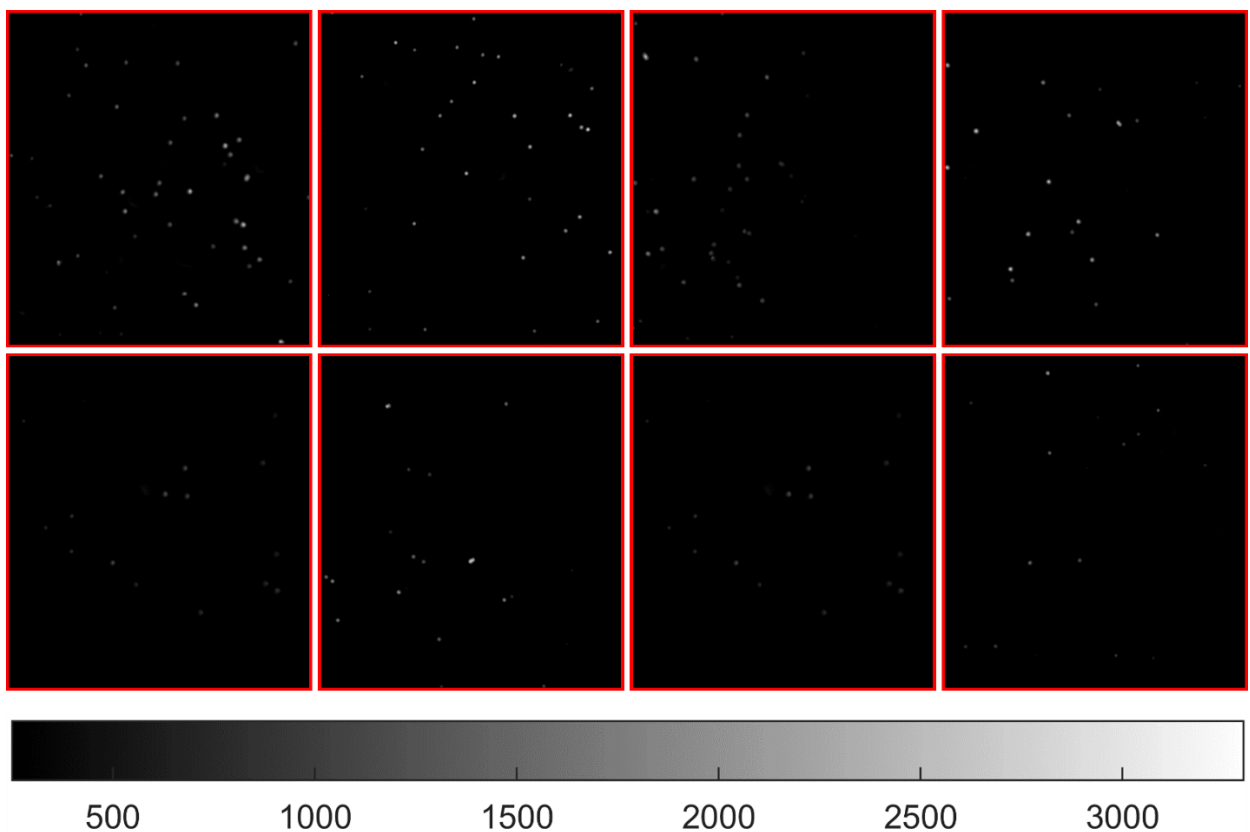

a

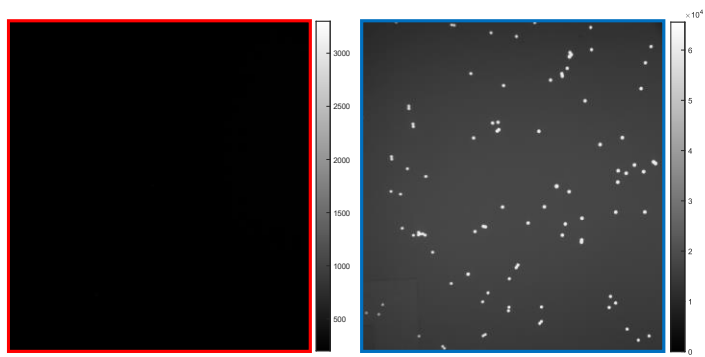

b

**Supplementary Figure 6. Fluorescence data for coupling and control reactions implemented in synthesis columns (benchtop reactions). a,** Beads with enzymatic coupling of fluorescently labelled (AF647) bases show a large variation in fluorescent intensity which indicates a large variation in the degree of coupling. Images were captured as 16-bit Tiff files using PCO.EDGE 5.5 camera with an integration time of 2s under red excitation ($\lambda_{ex} = 637nm$). A few representative snapshots are shown here. **b**, Control experiment implemented on beads without initiator strands exposed to coupling reagents without initiator strands shows no red fluorescence under red excitation ($\lambda_{ex} = 637nm$). The beads fluoresce green under blue excitation. This is used to detect the physical presence of the beads within the field of view of the microscope. The outline color of the images indicates the excitation source (blue $\lambda_{ex} - 455nm$, red $\lambda_{ex} - 637nm$).

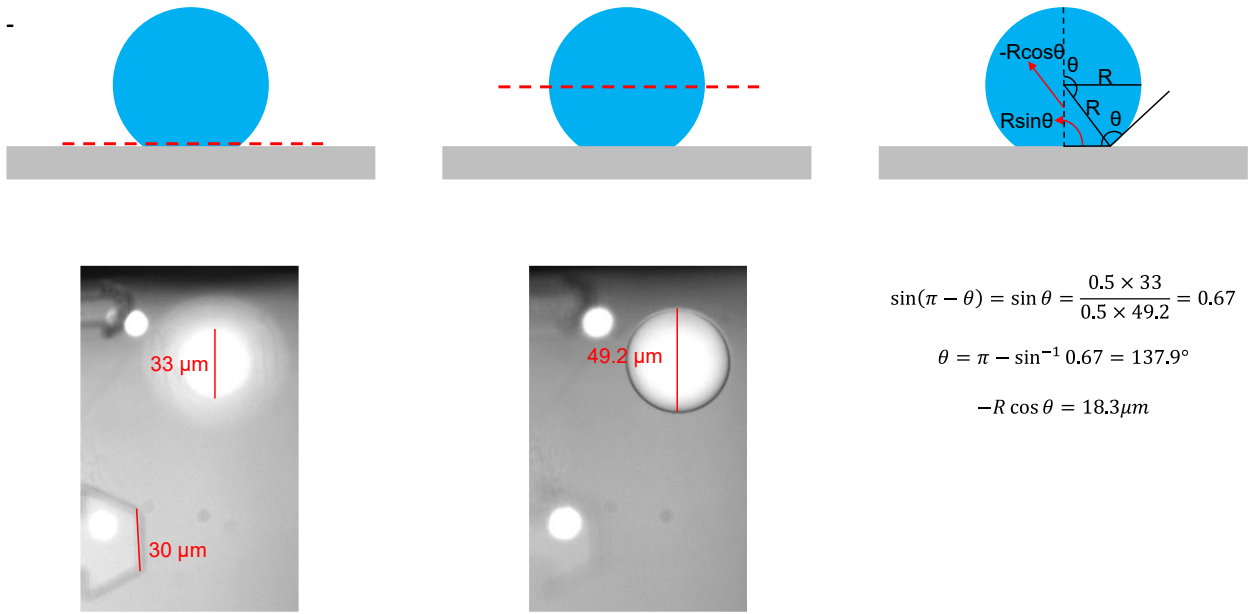

**Supplementary Figure 7. Optical microscopy-based estimation of the contact angle of the droplet on the inner walls of the device.** The contact angle is calculated using elementary trigonometric relationships from the estimates of the droplet diameter at the interface with the glass bottom of the device and at the greater circle of the droplet.

$$\sin(\pi - \theta) = \sin\theta = \frac{0.5 \times 33}{0.5 \times 49.2} = 0.67$$

$$\theta = \pi - \sin^{-1} 0.67 = 137.9°$$

$$-R\cos\theta = 18.3 \mu m$$

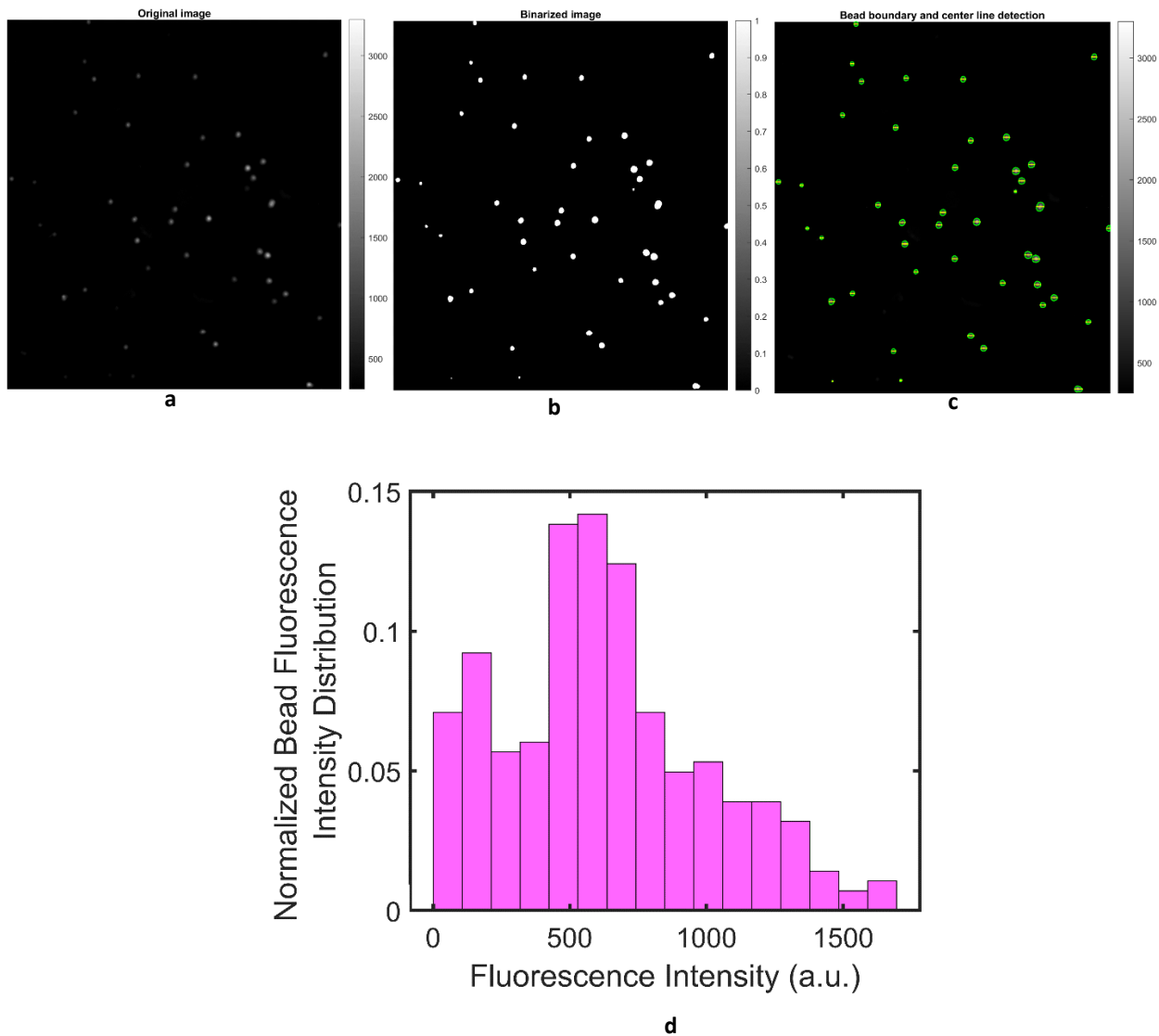

**Supplementary Figure 8. Image processing of benchtop reaction snapshots to extract fluorescence data from beads a,** A 16-bit snapshot of fluorescent beads post enzymatic coupling of fluorescently labelled nucleotides in benchtop setups under red excitation ($\lambda_{ex} = 637nm$). It is the same as the first snapshot in Fig. 6a. **b**, A binary transformation of the image in **a. c.** Edge and center line detection of the binarized images in **b**. The extracted fluorescence intensity patterns across the center lines are plotted in Fig. 5 of the main text. The transformations are applied to all the collected bead snapshots. **d**. Average fluorescence intensity distribution of all the beads across all the frames.

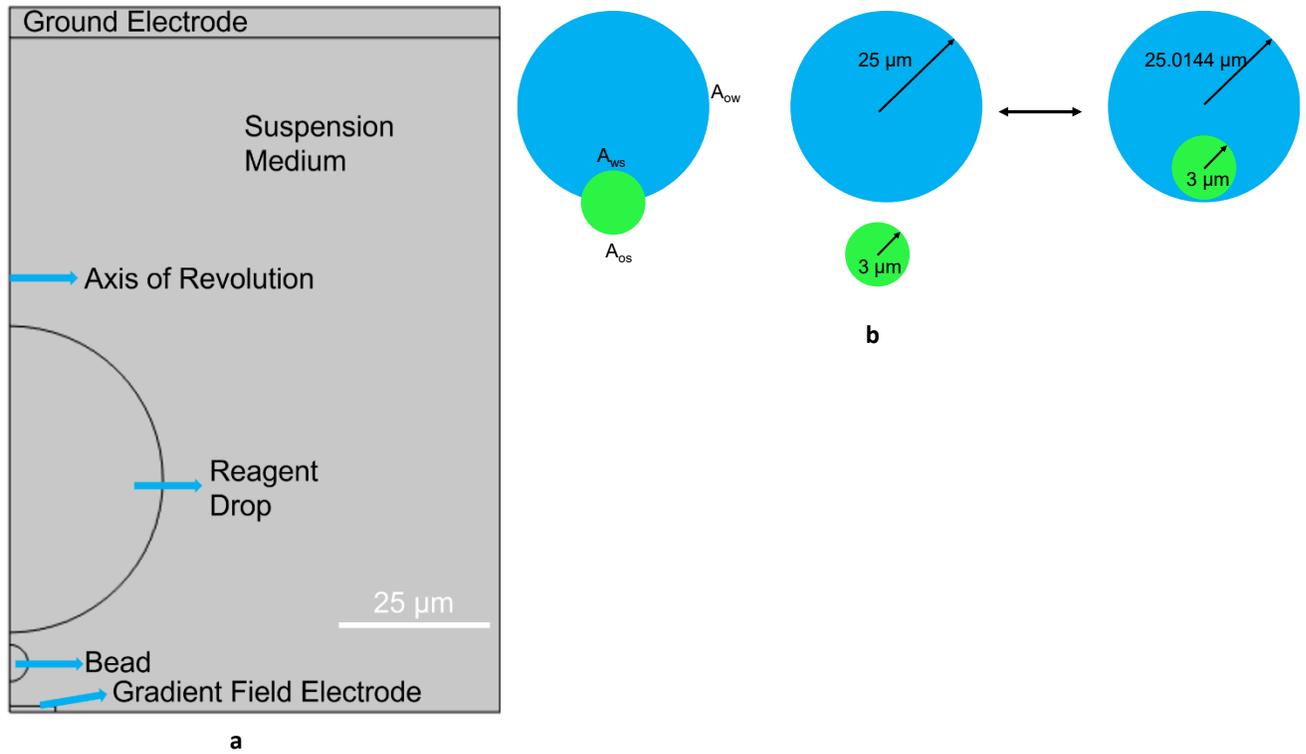

**Supplementary Figure 9. Electrohydrodynamic simulation setup and geometrical and experimental estimation of simulation parameters a,** Axis-symmetric simulation structure used for electrohydrodynamic simulations in COMSOL Multiphysics ®. **b**, Change in the dimension of the droplet with the encapsulation and ejection of the bead from the droplet. As can be seen the radius of the droplet and therefore $A_{ow}$ changes negligibly.

| Variable | Value | Description | Reference |
|---|---|---|---|
| $R_b$ | 3 µm | Radius of bead | 6µm diameter streptavidin coated green, fluorescent polystyrene beads purchased from polysciences (Catalog No.#-24157). |
| $R_d$ | 25 µm | Radius of drop | Estimated from measurements in Supplementary Information Fig. S1c |
| $R_{grad}$ | 15 µm | Radius of gradient field electrode | From designed dimension of the electrode Fig. 1(b) |
| $R_{ground}$ | 80 µm | Radius of ground electrode | Chosen to make the simulation space sufficiently larger than the dimensions of the bead and droplet under consideration. |
| $R_{medium}$ | 80 um (=$R_{ground}$) | Radius of suspension medium | Same as $R_{ground}$ due to the system geometry |
| $h_{grad}$ | 1 µm | Thickness of gradient field electrode | Approximate thickness of evaporatively deposited Indium Tin Oxide electrode which is 0.8µm. |
| $h_{ground}$ | 5 µm | Thickness of ground electrode | This choice does not affect the simulation in any way. |
| $h_{medium}$ | 115 µm | Thickness of medium | Spacing between the electrode and ground in Fig. 1(b) |
| $\rho_{drop}$ | 1025 kg/m$^3$ | Density of reagent phase | Evaluated by considering the density of all the ingredients of the reaction buffer. |
| $\rho_{SO}$ | 818 kg/m$^3$ | Density of oil | Density of silicone oil (CAS#-105-51-7) |
| $\rho_{Span80}$ | 986 kg/m$^3$ | Density of Span80 | Density of Span80 (CAS#-1338-43-8) |
| $V_{oil}$ | 200 ml | Volume of silicone oil | |
| $V_{Span80}$ | 4 ml | Volume of Span80 | |
| $V_{med}$ | $V_{oil} + V_{Span80}$ | Volume of medium | Approximate volume of oil + Span80 |
| $V_{of}$ | 200/204 | Volume fraction of oil | $V_{oil}/V_{med}$ |
| $V_{Span80f}$ | 4/204 | Volume fraction of Span80 | $V_{Span80f}/V_{med}$ |
| $\rho_{medium}$ | 821 kg/m$^3$ | Density of medium | $\rho_{SO} V_{of} + \rho_{Span80} V_{Span80f}$ |
| $\eta_{drop}$ | 8.9e-4 Pa.s | Dynamic viscosity of reagent drop | Assumed to be dynamic viscosity of water |
| $\eta_{SO}$ | 8.18e-4 Pa.s | Dynamic viscosity of 1cSt silicone oil | $= \kappa_{SO} \times \rho_{SO}$ ($\kappa_{SO}$ is the kinematic viscosity which is 1 cSt) |
| η$_{Span80}$ | 1 Pa.s | Dynamic Viscosity of Span80 | Viscosity of Span80 (CAS#-1338-43-8) |
| η$_{medium}$ | 1.5e-3 Pa.s | Dynamic viscosity of suspension medium | $\eta_{medium}^{1/3} = x_{SO}\eta_{SO}^{1/3} + x_{Span80}\eta_{Span80}^{1/3}$ [S4] |
| ε$_{drop}$ | 80 | Relative permittivity of drop | Relative permittivity of water |
| ε$_{medium}$ | 2.30 | Relative permittivity of medium | Relative permittivity of silicone oil (CAS#-105-51-7). |
| ε$_{bead}$ | 2.55 | Relative permittivity of bead | [S5] |
| σ$_{drop}$ | 6e-1 S/m | Electrical conductivity of reagent drops | Measured |
| σ$_{medium}$ | 1e-14 S/m | Electrical conductivity of suspension medium | [S6] |
| σ$_{bead}$ | 1e-9 S/m | Electrical conductivity of bead | Conductivity of polystyrene bead suspended in silicone oil [S7]. |
| γ | 5.5 mN/m | Interfacial tension between the reagent droplet and the suspension medium | Measured. Fig. 3 main text |
| θ$_{bead}$ | 145° | Contact angle of the reagent drop on the surface of the streptavidin coated bead surrounded by the oil medium. | Estimated Fig. 3 main text |
| θ$_{wall}$ | 140° | Contact angle of the reagent drop on the device walls. | Estimated Supplementary Information Fig. S1d |

**Supplementary Table 1.** Table of all the geometric and material parameters used in electrohydrodynamic simulations and energy calculations.

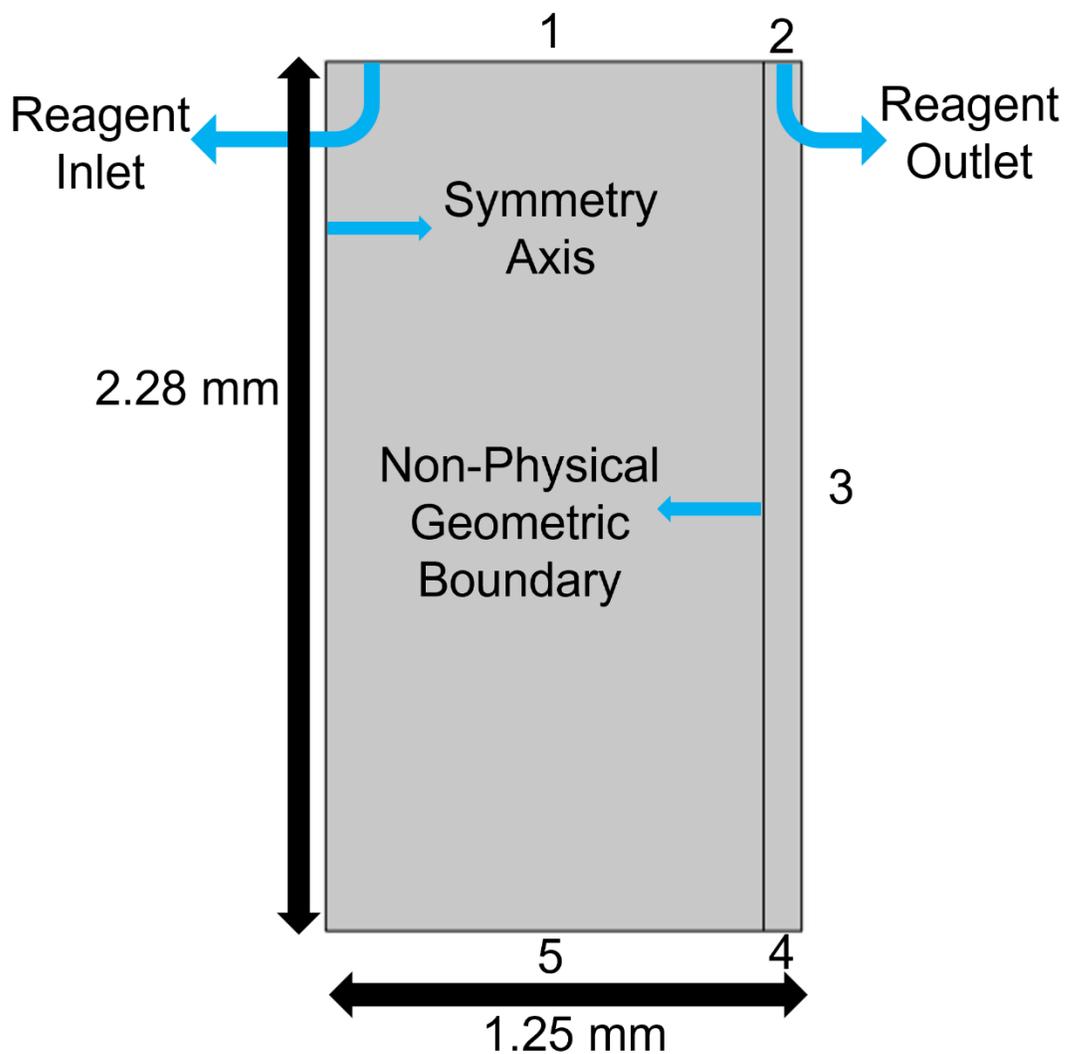

**Supplementary Figure 10. Turbulent flow driven particle tracking simulation setup.** Two-dimensional turbulent flow driven particle tracking simulation implemented in COMSOL Multiphysics® to study displacement patterns of particles from stacked configuration in synthesis columns as the reagent fluid is injected into the reaction space. 3-5 are walls for the fluid flow. 1-5 are walls for the particles as they are held in the space by filters. Reagent fluid enters from the top and tends to leak from the sides when the reaction space gets filled up. The filter at the bottom creates a resistive path for fluid flow vertically down, thus leading to its leakage from the sides.